\documentclass[aps,pre,twocolumn,preprintnumbers,floatfix,nofootinbib]{revtex4-1}

\usepackage{amsmath}
\usepackage{amssymb}

\usepackage{graphicx}
\usepackage{subcaption}
\usepackage{epstopdf}
\usepackage{dcolumn}
\usepackage{bm}
\usepackage{marvosym}
\usepackage{rotating}

\usepackage{color}
\definecolor{light-gray}{gray}{0.5}
\definecolor{blue}{rgb}{0.0,0.0,1.0}
\definecolor{green}{rgb}{0.0,0.5,0.0}
\definecolor{red}{rgb}{1.0,0.0,0.0}
\definecolor{cyan}{rgb}{0.0,0.75,0.75}
\definecolor{magenta}{rgb}{0.75,0.0,0.75}
\definecolor{yellow}{rgb}{0.75,0.75,0.0}

\newcommand{\avg}[1]{\langle{#1}\rangle}
\newcommand{\sdot}{\cdot}

\newcommand{\grad}{\bm \nabla}
\newcommand{\pd}{\partial}

\newcommand{\lr}[1]{\left(#1\right)}

\begin{document}
% \title{Symmetry breaking and universality of decaying magnetohydrodynamic Taylor-Green flows}
\title{Symmetry breaking and universality of decaying MHD Taylor-Green flows}
\author{V. Dallas}
\email{vassilios.dallas@lps.ens.fr}
\author{A. Alexakis}
\affiliation{Laboratoire de Physique Statistique, \'Ecole Normale Sup\'erieure, Universit\'e Pierre et Mari\'e Curie, Universit\'e Paris Diderot, CNRS, 24 rue Lhomond, 75005 Paris, France}

% It is not clear, however, if, when studying these flows at substantially higher Reynolds numbers, as was done in Ref. [15] for the Taylor-Green flows in magnetohydrodynamics but, contrary to Ref. [15], not imposing the symmetries at all time, one will still have three different scaling laws for the total energy spectra for these three configurations.

\begin{abstract}
We investigate the evolution and stability of a decaying magnetohydrodynamic (MHD) Taylor-Green flow.
The chosen flow has been shown to result in a steep total energy spectrum with power law behaviour
$k^{-2}$. We investigate the symmetry breaking of this flow by exciting perturbations of different amplitudes. It is shown that for any finite amplitude perturbation there is a high enough Reynolds number for which the perturbation will grow enough at the peak of dissipation rate resulting to a non-linear feedback in the flow and subsequently break the Taylor-Green symmetries. In particular, we show that symmetry breaking at large scales occurs if the amplitude of the perturbation is $\rho_{crit} \sim Re^{-1}$ and at small scales occurs if $\rho_{crit} \sim Re^{-3/2}$. 
%This symmetry breaking affects the scaling laws of the energy spectra towards a universal scaling regime for $Re \gg 1$, away from the $k^{-2}$ scaling, limiting the debate between the classical $k^{-5/3}$ and $k^{-3/2}$ power law energy spectra.
This symmetry breaking modifies the scaling laws of the energy spectra at the peak of dissipation rate away from the $k^{-2}$ scaling and towards the classical $k^{-5/3}$ and $k^{-3/2}$ power laws.
\end{abstract}

\maketitle

%%%%%%%%%%%%%%%%%%%%%%%%%%%%%%%%%%%%%%%%%%%%%%%%%%%
\section{\label{sec:intro}Introduction}
%%%%%%%%%%%%%%%%%%%%%%%%%%%%%%%%%%%%%%%%%%%%%%%%%%%
% Lack of knowledge of the precise power law scaling of the energy spectrum has implications to the prediction of many astrophysical phenomena, such as the heating rates of the solar corona and acceleration of the solar wind,1 the transport of mass and energy into the Earth’s magnetosphere.7, the dynamics of the interstellar medium, 2–4 the regulation of star formation,5 and the transport of heat in galaxy clusters,6. %see Howes-Nielson-Drake-Schroeder-Skiff-Kletzing-Carter_1306.1460v1.pdf

%The lack of prediction of heating rates in solar and space physics \cite{petersonfabian06,marinoetal08} is intimately connected to the slope of the energy spectrum. 

In magnetohydrodynamic (MHD) turbulence several phenomenological theories exist debating for the interpretation of the power law of the energy spectrum \cite{biskamp03,zhouetal04,boldyrev06,ngbhattacharjee97,galtieretal00}. In summary, the power law scaling exponents obtained in these phenomenologies based on weak and strong turbulence arguments both for isotropic and anisotropic energy spectra are $-2$, $-5/3$ and $-3/2$. Numerical simulations to date are unable to provide a definitive answer to this scaling. For example, some direct numerical simulations (DNS) obtained energy spectra with $k^{-5/3}$ while others $k^{-3/2}$ scaling for freely decaying MHD turbulent flows \cite{mullergrappin05,mininnipouquet07}. Astrophysical observations have shown that this difference in the power law scaling also exists for the measured energy spectra of the solar wind 
\cite{podestaetal07}. %The difference between $-5/3$ and $-3/2$ power laws is subtle enough (10\% difference) so that an inertial range of more than an order of magnitute is necessary to make a clear distinction between them. However, a $-2$ scaling exponent can be more transparent even for moderate Reynolds numbers, such as those obtained by DNS.
In addition, indications of $k^{-2}$ scaling are reported for the magnetic energy spectrum measured in the magnetosphere of Jupiter \cite{sauretal02}.

Recently, large resolution simulations by Lee et al. \cite{leeetal10} demonstrated $k^{-2}$, $k^{-5/3}$ and $k^{-3/2}$ total energy spectrum scalings for different initial conditions of the magnetic field. Thus, they showed dependence of the energy spectrum at the peak of dissipation on the initial conditions. Consequently, this suggests lack of universality in decaying MHD turbulence. The difference between $-5/3$ and $-3/2$ power laws is subtle enough (10\% difference) so that an inertial range of more than an order of magnitude is necessary to make a clear distinction between them. However, a $-2$ scaling exponent can be more transparent even for moderate Reynolds numbers, such as those obtained by DNS. For this reason, in this work we focus on the initial conditions the lead to the $k^{-2}$ spectrum. This scaling of the total energy spectrum was demonstrated to originate from high shearing regions that manifest discontinuities in the magnetic field corresponding to strong current sheets \cite{da13b}.

All the initial conditions in \cite{leeetal10} were satisfying symmetries of the Taylor-Green (TG) vortex \cite{brachetetal83}. This property was taken into account by numerically enforcing these symmetries in order to achieve higher resolutions with less compational cost \cite{leeetal08,leeetal10}.
%to gain substantial savings in both computing time and memory usage at a given Reynolds number. 
The $-2$ power law spectrum was also confirmed by Dallas \& Alexakis \cite{da13a,da13b} without imposing the TG symmetries, allowing thus the turbulence to evolve freely with the view that the initial TG vortex symmetries will break at high enough Reynolds numbers. However, even for their highest Taylor Reynolds number simulations ($\sim \mathcal O(100)$), the TG vortex symmetries did not break within the time interval of reaching the peak of dissipation. This suggests that the TG symmetries are a strong property of the evolution equations preserved in time. However, Stawarz et al. \cite{stawarzetal12} showed that the TG symmetries can be broken at very long time scales using runs of low Reynolds numbers %(but high enough for the flow to be turbulent) 
due to round-off error accumulation.

Preservation of the TG symmetries hinders the flow from exploring all phase space and concequently prevents it from reaching a universal behaviour. Moreover, the breaking of the TG symmetries can possibly modify the scaling of the energy spectrum by the time of maximum dissipation rate $t_{peak}$, where the largest inertial range is obtained. Thus, before claiming lack of universality of spectral exponents for decaying MHD turbulence in periodic boxes, the persistence of the TG symmetries within $t_{peak}$ is an important issue that needs to be resolved. %However, breaking the TG symmetries does not necessarily imply return to universality. \textcolor{red}{It is also possible that even if the symmetries break the cascade mechanism does not alter and the slope of the energy spectrum, for example, remains the same}.

We expect a critical perturbation amplitude to exist so that the system transitions 
from symmetry preservation to symmetry breaking within $t_{peak}$. The dependence of this critical amplitude on the Reynolds number and whether the breaking of the TG symmetries leads to a different spectral exponent are the key open questions that we address in this work. In order to demonstrate lack of universality at the peak of dissipation one needs to show if, at $Re \gg 1$, there is a finite perturbation amplitude below which the power law of the spectrum remains unchanged. Showing this way that the set of initial conditions which lead to a specific behaviour is of non-zero measure in the limit of $Re \rightarrow \infty$.

In summary, given an infinitesimal perturbation, is there a Reynolds number such that the symmetries break within $t_{peak}$? Will the breaking of the symmetries lead to a different power law spectrum? Are the discontinuities, which are responsible for the $k^{-2}$ spectra, formed due to enforcement of the TG symmetries? Are there universality classes for moderate Reynolds numbers or is there a universal power law scaling for the high Reynolds number limit? %These are the questions we address in this paper.
In this work, we investigate these questions by considering a large set of numerical simulations.

The paper is structured as follows. Section \ref{sec:dns} describes the numerical methodology to solve the governing equations for our decaying MHD turbulent flows and section \ref{sec:TG} provides the necessary details with regards to the Taylor-Green vortex, its symmetries and the measures of symmetry breaking. In section \ref{sec:param}, we define our numerical parameters along with our perturbed initial conditions. First, we analyse the results from the growth of infinitesimal perturbations (see section \ref{sec:smallpert}) and then from the finite amplitude perturbations (see section \ref{sec:largepert}) by applying the measures of symmetry breaking. Finally, in section \ref{sec:end} we conclude by summarising our findings.

% %%%%%%%%%%%%%%%%%%%%%%%%%%%%%%%%%%%%%%%%%%%%%%%%%%%
% The TG symmetries are preserved exactly in the absence of any perturbation or
% any external noise. If however such a small perturbation exists (whether artificial
% as numerical noise or as an input in the initial conditions) it can either decay or 
% grow. For the decaying MHD flows that we examine here, growth of the perturbation is not sufficient for the 
% symmetries to be broken. It also needs to grow fast enough so that the perturbation 
% becomes non-linear by the time the peak of dissipation has been reached.  
% %%%%
% 
% %%%%%%%%%%%%%%%%%%%%%%%%%%%%%%%%%%%%%%%%%%%%%%%%%%%

%%%%%%%%%%%%%%%%%%%%%%%%%%%%%%%%%%%%%%%%%%%%%%%%%%%
\section{\label{sec:dns}DNS of decaying MHD turbulence}
%%%%%%%%%%%%%%%%%%%%%%%%%%%%%%%%%%%%%%%%%%%%%%%%%%%
We consider the three-dimensional, incompressible MHD equations of fluid velocity $\bm u$ and magnetic induction $\bm b$ to be
\begin{align}
 \pd_t \bm u - (\bm u \times \bm \omega) &= - \grad P + \nu \bm\Delta \bm u + (\bm j \times \bm b)
 \label{eq:ns} \\
 \pd_t \bm b + (\bm u \sdot \grad) \bm b &=  (\bm b \sdot \grad) \bm u + \kappa \bm\Delta \bm b
 \label{eq:induction} \\
 \grad \sdot \bm u &= \grad \sdot \bm b = 0
 \label{eq:incomp}
\end{align}
with $\nu$ the kinematic viscosity, $\kappa$ the magnetic diffusivity,  $\bm \omega \equiv \grad \times \bm u$ the vorticity, $\bm j \equiv \grad \times \bm b$ the current density and $P = p/\rho + \tfrac{1}{2}\bm u^2$ the fluid pressure, composed by the plasma pressure $p$ divided by $\rho$ the constant mass density plus the hydrodynamic pressure $\tfrac{1}{2}\bm u^2$. Note that the magnetic field has units of Alfv\'en velocity, i.e. $\bm b/\sqrt{\rho \mu_0}$, where $\mu_0 = (\kappa \sigma)^{-1}$ is the permeability of free space with $\sigma$ the electrical conductivity. If $\nu = \kappa = 0$, the total energy $E_t \equiv \frac{1}{2}\avg{|\bm u|^2 + |\bm b|^2} = E_u + E_b$, the magnetic helicity $H_b \equiv \avg{\bm u \sdot \bm b}$ and the cross helicity $H_c \equiv \avg{\bm a \sdot \bm b}$ are conserved in time (the angle brackets $\avg{.}$ denote spatial averages in this study). Here, $\bm a$ is the magnetic potential, which is defined as $\bm a \equiv - \grad^{-2}(\grad \times \bm b)$, since one can set 
$\bm b \equiv \grad \times \bm a$ with $\grad \sdot \bm a = 0$. 

Our numerical method is pseudo-spectral \cite{gottlieborszag77}, where each component of $\bm u$ and $\bm b$ is represented as truncated Galerkin expansions in terms of the Fourier basis. The non-linear terms are initially computed in physical space and then transformed to spectral space using fast Fourier transforms \cite{fftw98}. Aliasing errors are removed using the 2/3 dealiasing rule, i.e. wavenumbers $k \in [1,N/3]$, where $N$ is the number of grid points in each Cartesian coordinate of our box of period $2\pi$. The non-linear terms along with the pressure term are computed in such a way that $\bm u$ and $\bm b$ are projected on to a divergence-free space so that Eqs. \eqref{eq:incomp} are satisfied \cite{mpicode05a}. The temporal integration of Eqs. \eqref{eq:ns} and \eqref{eq:induction} is performed using a second-order Runge-Kutta method. The code is parallelised using a hybrid parallelisation (MPI-OpenMP) scheme \cite{hybridcode11}.

%%%%%%%%%%%%%%%%%%%%%%%%%%%%%%%%%%%%%%%%%%%%%%%%%%%
\section{\label{sec:TG}Taylor-Green vortex, symmetries and  measures of symmetry breaking}
%%%%%%%%%%%%%%%%%%%%%%%%%%%%%%%%%%%%%%%%%%%%%%%%%%%
The initial conditions that we choose to focus in this study is a magnetic Taylor-Green flow, which results in $k^{-2}$ spectra at the peak of dissipation \cite{leeetal10,pouquetetal10,da13a,da13b}. In particular, the initial velocity field is the Taylor-Green vortex \cite{taylorgreen37} defined as 
\begin{equation}
 \label{eq:TGu}
 \bm u_{TG}(\bm x) = u_0 %(\sin x \cos y \cos z, -\cos x \sin y \cos z, 0)
  \begin{pmatrix}
  & \sin x \cos y \cos z \\
 -& \cos x \sin y \cos z \\
  &          0
 \end{pmatrix},
\end{equation}
and the initial magnetic field is given by 
\begin{equation}
 \label{eq:TGb}
 \bm b_{TG}(\bm x) = b_0 %(\cos x \sin y \sin z, \sin x \cos y \sin z, -2 \sin x \sin y \cos z)
 \begin{pmatrix}
  \cos x \sin y \sin z \\
  \sin x \cos y \sin z \\
  -2 \sin x \sin y \cos z
 \end{pmatrix}
\end{equation}
where $b_0$ and $u_0$ were chosen so that the 
norm of the two fields is unity, i.e. $\| \bm u_{TG} \|=\| \bm b_{TG} \|=1$. Here $\| . \|$ stands for the $L_2$ norm $\| \bm g \|^2 = \frac{1}{V} \int_V \bm g \cdot \bm g \,d^3x$, where $\bm g$ is an arbitrary vector field.
%satisfying the following relations $\bm b_{TG} = -(b_0/u_0) \grad \times \bm u_{TG}$ and $\bm u_{TG} = (u_0/b_0) \grad \times \bm b_{TG}$.

Given these initial conditions and in the absence of any noise the symmetries are preserved by the evolution equations exactly 
\cite{leeetal08,da13a}. In particular, we have 
reflection (anti)symmetries about the planes $x=0$, $x=\pi$, $y=0$, $y=\pi$, $z=0$ and $z=\pi$ 
as well as rotational (anti)symmetries of angle $n\pi$ about the axes $(x,y,z)=(\tfrac{\pi}{2},y,\tfrac{\pi}{2})$ 
and 
$(x,\tfrac{\pi}{2},\tfrac{\pi}{2})$ and of angle $n\pi/2$ about the axis $(\tfrac{\pi}{2},\tfrac{\pi}{2},z)$ for $n \in \mathbb{Z}$.
The above mentioned planes that possess reflection symmetries form the insulating faces of the sub-boxes $[0,\pi]^3$ 
\cite{brachetetal83}, where the $\bm j_{TG}$ is everywhere parallel to these faces. Note that for these particular 
initial conditions $\bm b_{TG}$ satisfies the same symmetries with $\bm \omega_{TG}$ and $\bm u_{TG}$ with $\bm j_{TG}$. 
%Moreover, the magnetic and cross helicity are globally restricted to vanish for all times due to these symmetries.

It was shown in \cite{da13b} that the $k^{-2}$ spectrum observed in the numerical simulations originates from the formation of strong current sheets at the reflection symmetry planes $x=0$, $x=\pi$, $y=0$ and $y=\pi$. So, we focus on only one of these symmetries. In particular, we will investigate the reflection symmetry around the plane $x=0$. 
We then define the reflection operator $\bm R_x$ around the $x=0$ plane as 
\begin{equation}
\bm R_x   
\begin{pmatrix}
   g_x (x,y,z) \\
   g_y (x,y,z) \\
   g_z (x,y,z)
 \end{pmatrix}
=
\begin{pmatrix}
  -g_x (-x,y,z) \\
   g_y (-x,y,z) \\
   g_z (-x,y,z)
\end{pmatrix}.
\end{equation}
The TG initial conditions under the action of $\bm R_x$ transform as follows
\begin{equation}
\bm R_x \bm u_{TG} = \bm u_{TG}  \quad \mathrm{and} \quad \bm R_x \bm b_{TG} = -\bm b_{TG}.
\label{eq:reflection}
\end{equation}
Given any arbitrary set of fields $\bm u, \bm b$ we define $\bm u_s$ and $\bm b_s$ as
\begin{equation}
\bm u_s = \frac{1}{2} (\bm u + \bm R_x \bm u)
\quad \mathrm{and} \quad
\bm b_s = \frac{1}{2} (\bm b - \bm R_x \bm b)
\end{equation}
with $\bm u_s$ and $\bm b_s$ transforming similar to the TG initial conditions under reflection $\bm R_x$ 
(see Eq. \eqref{eq:reflection}).
Similarly we define $\bm u_a$ and $\bm b_a$
\begin{equation}
\bm u_a = \frac{1}{2} (\bm u - \bm R_x \bm u)
\quad \mathrm{and} \quad
\bm b_a = \frac{1}{2} (\bm b + \bm R_x \bm b)
\end{equation}
as the part of the flow that does not follow the TG symmetries. 
Note that $\bm u_a$ and $\bm b_a$ transform differently under reflection, i.e.
\begin{equation}
\bm R_x \bm u_a = -\bm u_a \quad \mathrm{and} \quad \bm R_x \bm b_a = \bm b_a.
\end{equation}
We will refer to $\bm u_s, \bm b_s$ as the symmetric part of the flow 
while to $\bm u_a, \bm b_a$ as the asymmetric part of the flow.
Note that if we start with $\bm u = \bm u_{TG}$ and $\bm b = \bm b_{TG}$ at $t=0$, then
$\bm u_a, \bm b_a$ will remain zero throughout the computation. Thus, $\bm u_a, \bm b_a$ can provide us with a measure of the extent the symmetries are broken.
Here we will focus on two such measures. First we consider the ratio of the energies
of asymmetric to the symmetric component of the fields $E_a/E_s$, where
\begin{equation}
E_a = \frac{1}{2} ( \| \bm u_a \|^2 + \| \bm b_a \|^2 )
%\quad \mathrm{and} \quad
\end{equation}
and
\begin{equation}
 E_s = \frac{1}{2} ( \| \bm u_s \|^2 + \| \bm b_s \|^2 ).
\end{equation}
%$E_a = \frac{1}{2} ( \| \bm u_a \|^2 + \| \bm b_a \|^2 )$  and 
%$E_s = \frac{1}{2} ( \| \bm u_s \|^2 + \| \bm b_s \|^2 )$.
This quantity provides a measure of the degree the TG symmetries are broken 
in the large (energy containing) scales. We also focus on the small
scales by looking at the ratio of the dissipation rates $\epsilon_a/\epsilon_s$, where
\begin{equation}
 \epsilon_a = \nu \| \grad \times \bm u_a\|^2 + \kappa \| \grad \times \bm b_a\|^2
\end{equation}
and
\begin{equation}
 \epsilon_s = \nu \| \grad \times \bm u_s\|^2 + \kappa \| \grad \times \bm b_s\|^2.
\end{equation}
%$\epsilon_a = \nu (\| \grad \times \bm u_a\|^2 + \| \grad \times \bm b_a\|^2)$ and
%$\epsilon_s = \nu (\| \grad \times \bm u_s\|^2 + \| \grad \times \bm b_s\|^2)$.

%%%%%%%%%%%%%%%%%%%%%%%%%%%%%%%%%%%%%%%%%%%%%%%%%%
\section{\label{sec:param}Initial conditions and simulation parameters}
%%%%%%%%%%%%%%%%%%%%%%%%%%%%%%%%%%%%%%%%%%%%%%%%%%
To study the stability of the TG symmetries and their implications 
on the energy spectrum a series of numerical simulations were performed.
The simulations were carried out on a triple periodic box of size $2\pi$. 
The initial conditions were composed by the TG initial conditions plus small perturbation fields 
$\sqrt{\rho}\, \bm u_p, \sqrt{\rho}\, \bm b_p$, viz. 
\begin{equation}
 \bm u = \bm u_{TG} + \sqrt{\rho}\, \bm u_p \quad \mathrm{and} \quad
 \bm b = \bm b_{TG} + \sqrt{\rho}\, \bm b_p.
\end{equation}
The perturbation fields $\bm u_p, \bm b_p$ 
were chosen to be a superposition of Fourier modes in spherical shells $2\le | \bm k | \le k_p$.
The phases of the Fourier modes were chosen so that $\bm R_x \bm u_p = -\bm u_p$, $\bm R_x \bm b_p = \bm b_p$ and random otherwise. In this way we guarantee that the two perturbation fields give no contribution to $\bm u_s$ and $\bm b_s$. 
The norm of the two fields was set to unity $\| \bm u_p \|=\| \bm b_p \|=1$ so that
the amplitude $\rho$ at $t=0$ expresses the ratio of the kinetic energy of the perturbation field to the energy of the TG flow (i.e. $\rho \equiv E_a|_{t=0}/E_s|_{t=0}$). Additionally, we define
$\rho_\epsilon \equiv \epsilon_a|_{t=0}/\epsilon_s|_{t=0}$ as the ratio of the dissipation rate 
of the asymmetric part of the flow to the symmetric part of the flow at $t=0$.

The Reynolds number is defined based on the velocity rms value at $t=0$ and smallest 
wavenumber $k_{TG}=1$ in the box, i.e. $Re \equiv \| u_{TG} \| / \nu k_{TG}$. With these scales we can also define the eddy turnover time $\tau_L \equiv (u_{TG}k_{TG})^{-1} = 1$ at $t=0$. The smallest length scale in our flows is defined based on Kolmogorov scaling $\eta \equiv (\nu^3 / \epsilon_t)^{1/4}$, where $\epsilon_t = \nu \avg{|\bm \omega|^2} + \kappa \avg{|\bm j|^2}$ is the total dissipation rate of energy. In all runs $\nu = \kappa$ and thus the Prandtl number is always unity. The set of parameters for all the examined runs is given in Table \ref{tbl:dnsparam}. 

\begin{table}[!ht]
  \caption{Numerical parameters of the DNS. For all runs $\nu=\kappa$. Note that $k_{max} = N/3$, using the $2/3$ dealiasing rule and the values of $k_{max}\eta$ are reported at the peak of $\epsilon_t$. }
  \label{tbl:dnsparam}
%\resizebox{0.5\textwidth}{!}
{
     \begin{tabular}{| c c c | c c c | c c c |}
     \hline
     \multicolumn{3}{ |c| }{$\rho=10^{-6}$, $k_p=10$ } &
     \multicolumn{3}{ c| }{$\rho=0.01$, $k_p=4$  } &
     \multicolumn{3}{ c| }{$\rho=0.1$, $k_p=4$  } \\
     \multicolumn{3}{ |c| }{$\rho_\epsilon =1.8 \sdot 10^{-5}$ \qquad } &
     \multicolumn{3}{ c| }{$\rho_\epsilon =0.054$ \qquad} &
     \multicolumn{3}{ c| }{$\rho_\epsilon =0.54$ \qquad }\\
     \hline
	  $Re$ &  $N$ & $k_{max}\eta$ & $Re$ &  $N$ & $k_{max}\eta$ & $Re$ & $N$  & $k_{max}\eta$ \\
     \hline
      50  &  128 & 2.80 &   50 &  128 & 2.80 &   50 &  128 & 2.78 \\
      100 &  128 & 1.69 &  100 &  128 & 1.69 &  100 &  128 & 1.67 \\
      200 &  256 & 2.09 &  200 &  256 & 2.09 &  200 &  256 & 2.04 \\
      300 &  256 & 1.48 &  300 &  256 & 1.47 &  300 &  256 & 1.43 \\
      500 &  512 & 2.25 &  500 &  512 & 2.24 &  500 &  512 & 2.15 \\
     1000 &  512 & 1.42 & 1000 &  512 & 1.42 & 1000 &  512 & 1.34 \\
     2000 & 1024 & 1.81 & 2000 & 1024 & 1.79 & 2000 & 1024 & 1.66 \\
          &      &      & 5000 & 2048 & 1.96 &      &      &      \\
     \hline
    \end{tabular}
}
%   \end{ruledtabular}
\end{table}

%%%%%%%%%%%%%%%%%%%%%%%%%%%%%%%%%%%%%%%%%%%%%%%%%%%
\section{\label{sec:smallpert}Growth of infinitesimal perturbations}
%%%%%%%%%%%%%%%%%%%%%%%%%%%%%%%%%%%%%%%%%%%%%%%%%%%
As a first step we look at the temporal evolution of flows with energy ratio 
$\rho=10^{-6}$ and dissipation ratio $\rho_\epsilon =1.8 \sdot 10^{-5}$ at $t=0$.
For this choice and for all Reynolds numbers 
considered here the amplitude of the perturbation 
(symmetry breaking part of the flow) remains 
much smaller than the symmetric part of the flow at all times of interest.
Thus, there is negligible effect of the perturbation on the part 
of the flow that obeys the TG symmetries and $\bm u_a$,  $\bm b_a$ 
evolve passively following the MHD equations \eqref{eq:ns} \& \eqref{eq:induction} linearised 
around $\bm u_s, \bm b_s$.

\begin{figure}[!ht]
  \includegraphics[width=8cm]{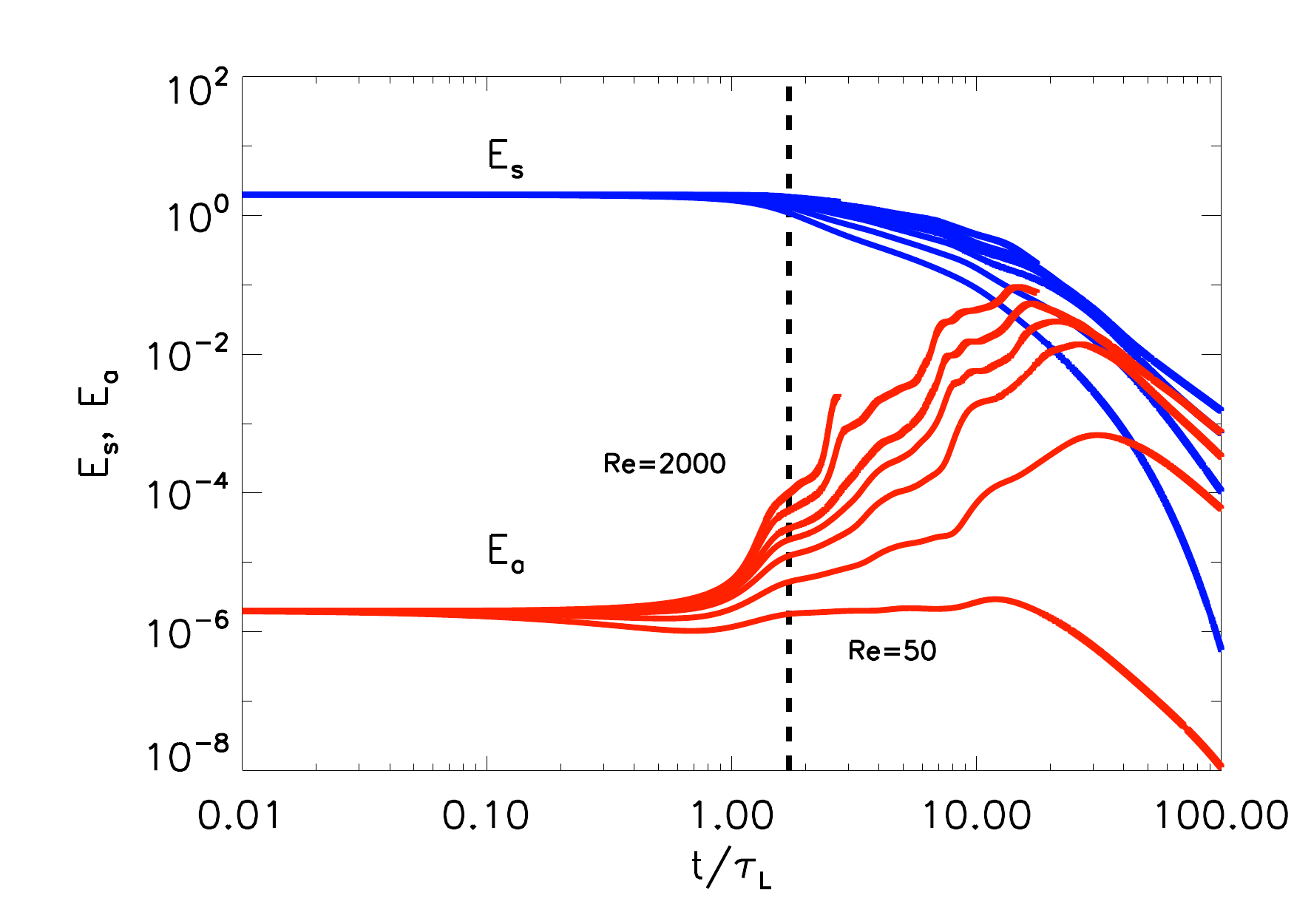}
  \caption{(Color online) Evolution of $E_s$ and $E_a$ as a function of time for different Reynolds numbers. The vertical dashed line indicates the time of maximum total dissipation rate.}
  \label{fig:EnR6}
 \end{figure}

Figure \ref{fig:EnR6} shows the temporal evolution of $E_s$ in blue (dark grey) and $E_a$ 
in red (light grey) for the seven different Reynolds numbers examined. The lower curves
are the small $Re$ cases while the top curves are the high $Re$ cases.
The vertical dashed line indicates $t_{peak}$, the time that $\epsilon_t$ is peaked which is the time we are interested in. 
In this case, very weak variations of $t_{peak}$ were observed with $Re$. It is evident that as 
the Reynolds number is increased the growth of the asymmetric part of
the energy $E_a$ is increased. 

\begin{figure}[!ht]
  \includegraphics[width=8cm]{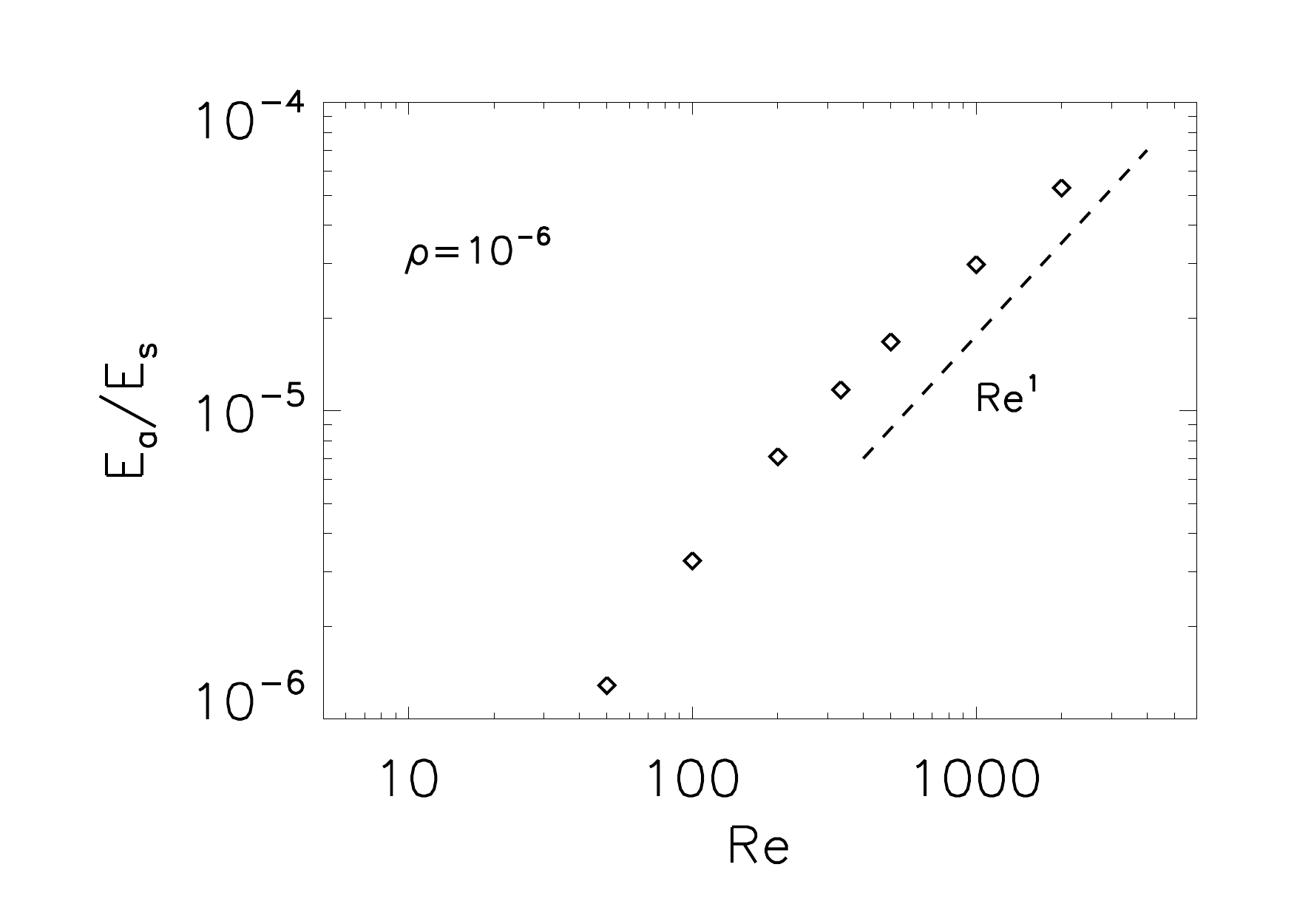}
  \caption{Energy ratio $E_p/E_s$ at the time of maximum dissipation rate as function of the Reynolds number for $\rho=10^{-6}$.}
  \label{fig:Ener_rat6}
 \end{figure}

Since we are interested in the symmetry breaking at the peak of dissipation, we plot the ratio $E_a/E_s$ at $t_{peak}$ as a function of the Reynolds number (see Fig. \ref{fig:Ener_rat6}).
This energy ratio appears to increase linearly with the Reynolds number. 
This linear increase of the perturbation energy can be understood if
we consider that the main source of growth of the perturbation comes
from the magnetic shear layer at the $x=0$ plane whose strength increases
with Re. The time scale for the growth of a perturbation in such layers is controlled 
by the shear rate $B_0/\delta$ where $B_0$ is the amplitude of the 
magnetic field in the layer and $\delta$ is the thickness of the shear layer. 
$B_0$ has negligible dependence on $Re$ and is determined  by the initial conditions.
The thickness of the reconnection layer $\delta$ is expected to scale like
$\delta \sim L/\sqrt{S_L}$ where $L$ is the length of the layer that is of the order of the box size 
and $S_L$ the Lundquist number $S_L = B_0 L / \kappa$ \cite{parker94}. 
For this problem $S_L \sim Re$ since $\| \bm u \| \sim \| \bm b \|$ and $\nu = \kappa$.
At the short time scale $t_{peak}$ we expect that transient growth rates will dominate and the growth of the perturbation in time will be linear rather than exponential. Therefore, we expect that the amplitude of the perturbation $A_p$
at $t_{peak}$ will increase from the initial value $A_0$ as: 
 \begin{eqnarray} 
  A_p & \sim & A_0 \frac{B_0}{\delta} t_{peak} \nonumber \\
    & \sim & A_0 B_0 \frac{ t_{peak} }{L} Re^{1/2} 
 \end{eqnarray}
from which we conclude that
 \begin{equation}
  \frac{E_a}{E_s} \sim \frac{A_p^2}{B_0^2} 
                          \sim A_0^2 \lr{\frac{t_{peak}}{L}}^2 Re
 \end{equation}
and hence the linear increase observed in Fig. \ref{fig:Ener_rat6}.

\begin{figure}[!ht]
  \includegraphics[width=8cm]{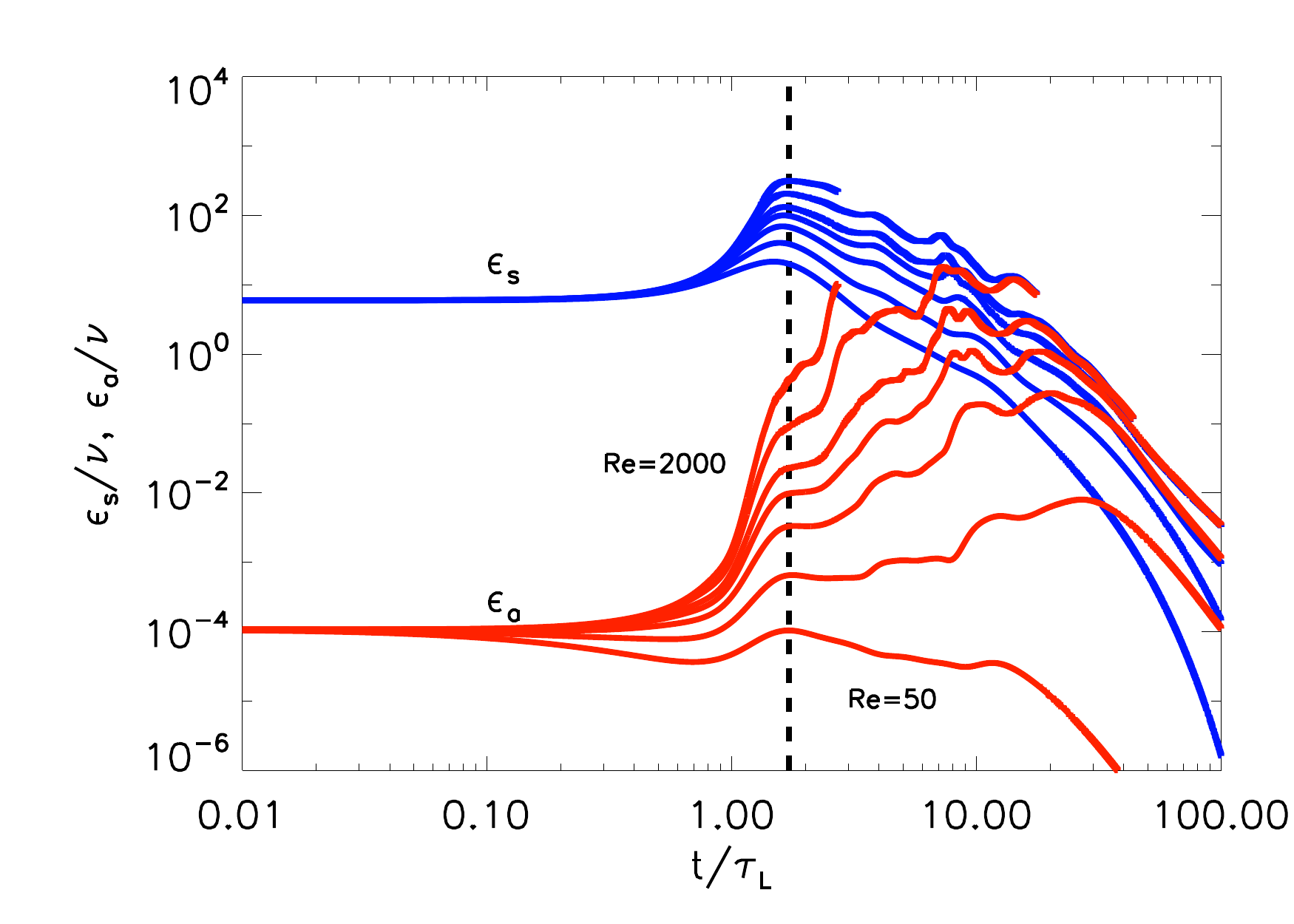}
  \caption{(Color online) Evolution of $\epsilon_s/\nu$ and $\epsilon_a/\nu$ as a function of time for different Reynolds numbers. The vertical dashed line indicates the time of maximum total dissipation rate.}
  \label{fig:EsR6}
 \end{figure}

Figure \ref{fig:EsR6} presents the time evolution of the two 
dissipations $\epsilon_s$ and $\epsilon_a$ divided by the viscosity $\nu$. 
While both grow with time, the asymmetric dissipation $\epsilon_a/\nu$ increases by roughly four orders of magnitude in two turnover times for the highest Reynolds number examined.

\begin{figure}[!ht]
  \includegraphics[width=8cm]{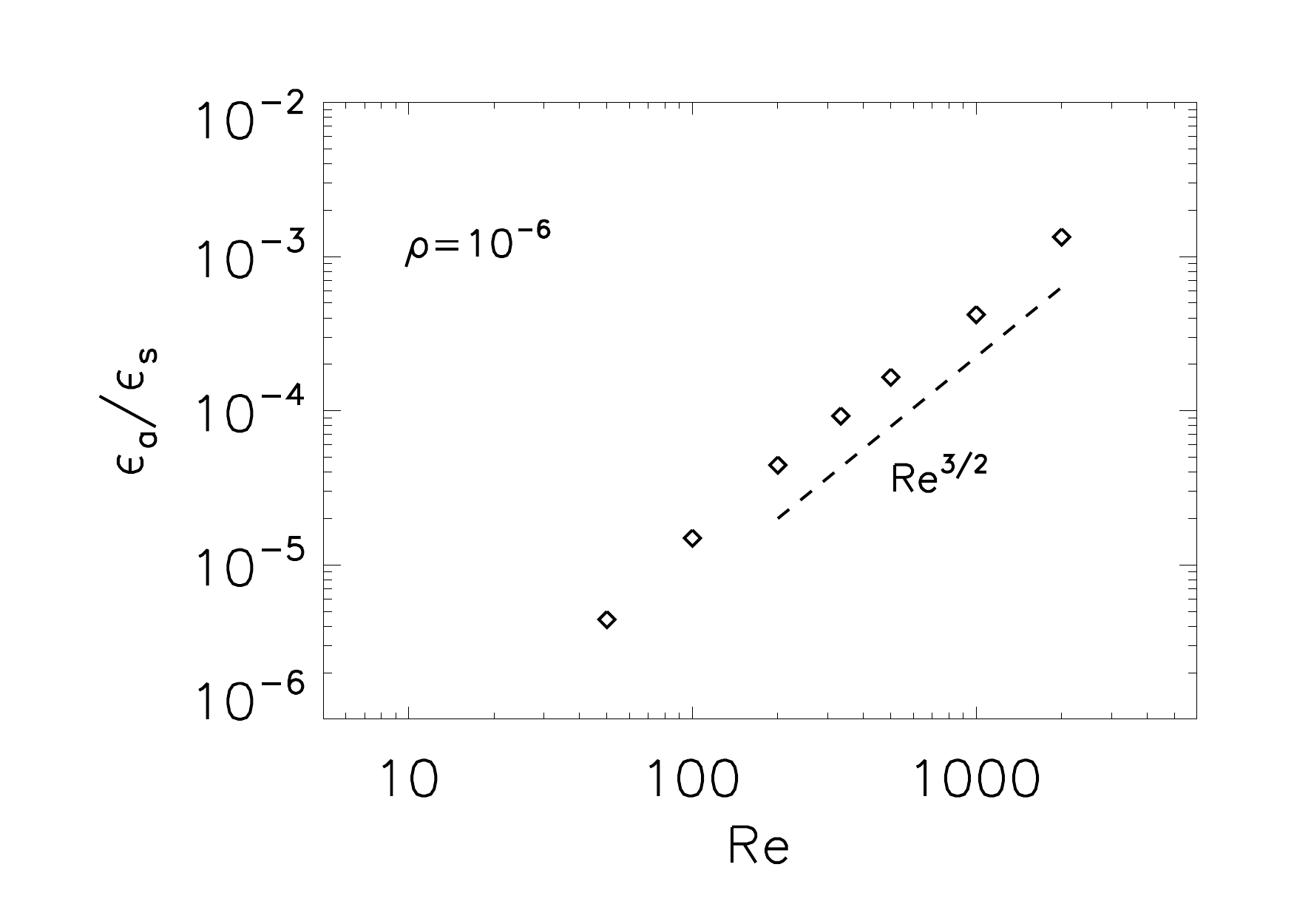}
  \caption{Dissipation ratio $\epsilon_p/\epsilon_s$ at the time of maximum dissipation rate as function of the Reynolds number for $\rho=10^{-6}$.}
  \label{fig:Enst_rat6}
 \end{figure}

In Fig. \ref{fig:Enst_rat6} we plot the ratio of the two dissipations $\epsilon_a / \epsilon_s$ at
$t_{peak}$. %An even steeper increase is observed for the dissipation ratio $\epsilon_a/\epsilon_s$.
This ratio is increasing faster than linear with Reynolds number, indicating that symmetries break even faster in the small scales. The scaling observed is close to
\begin{equation}
\epsilon_a / \epsilon_s \sim Re^{3/2}. %it would be nice to have an argument
\end{equation}

\begin{figure}[!ht]
   \begin{subfigure}{8cm}
   \includegraphics[width=\textwidth]{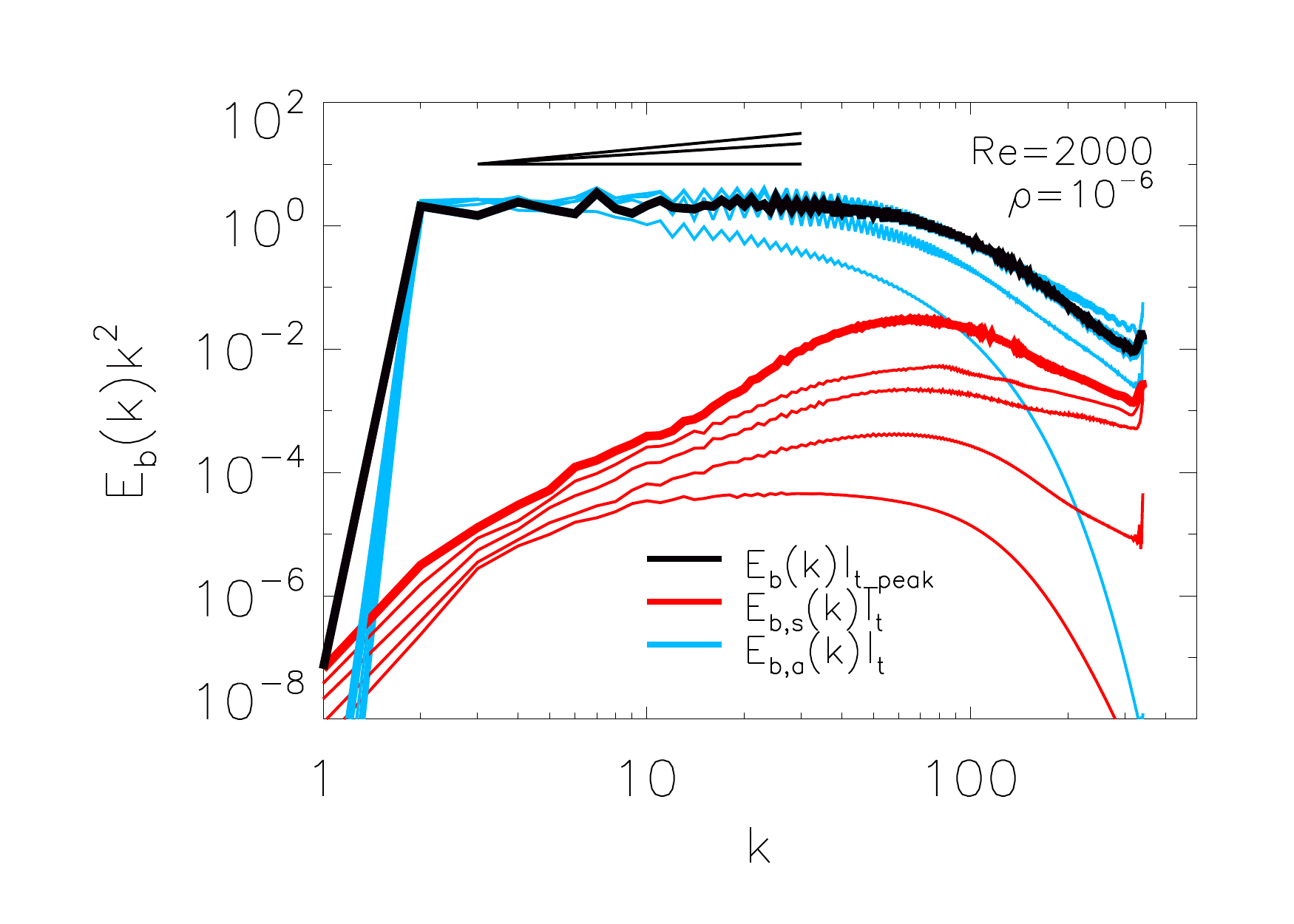}
   \caption{}
  \end{subfigure}
   \begin{subfigure}{8cm}
   \includegraphics[width=\textwidth]{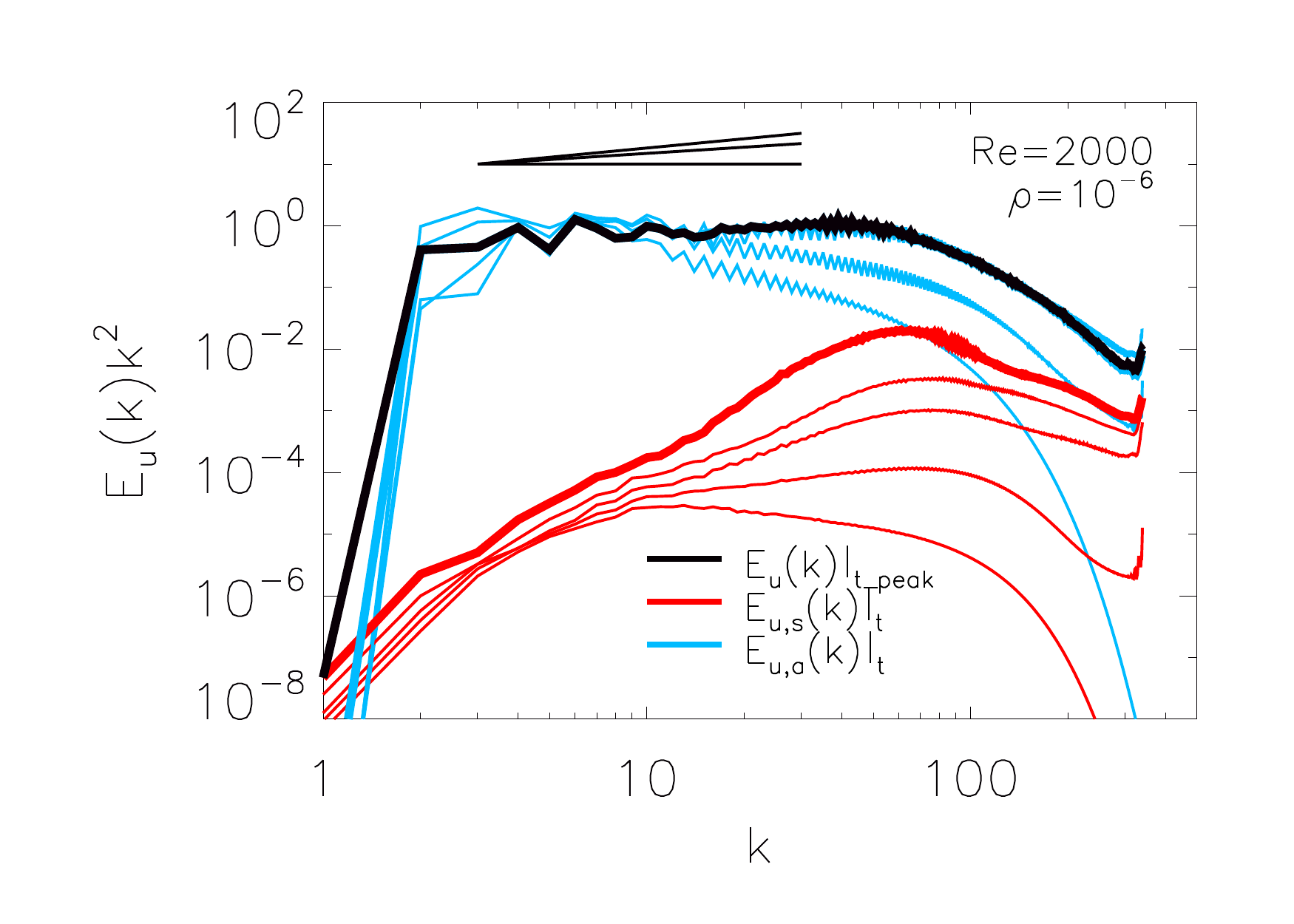}
   \caption{}
  \end{subfigure}
  \caption{(Color online) (a) Magnetic energy and (b) kinetic energy spectra at different times compensated by $k^2$ for $\rho=10^{-6}$ and $Re=2000$. The straight lines indicate the proposed spectral slopes $k^{-2},k^{-5/3},k^{-3/2}$.}
  \label{fig:Espec6}
 \end{figure}

The fast breaking of the symmetries in the small scales can also be seen by looking at the
energy spectra of the fields $\bm u_s, \bm u_a, \bm b_s$ and $\bm b_a$. 
Figure \ref{fig:Espec6}a and \ref{fig:Espec6}b show the symmetric and asymmetric part of the magnetic ($E_{b,s}, E_{b,a}$) and the kinetic ($E_{u,s}, E_{u,a}$) energy spectra, respectively,
compensated by $k^2$ for different times up to $t_{peak}$. Red (dark grey) curves represent the energy spectra of the asymmetric part of the flow, while blue curves (light grey) the energy spectra of the symmetric part of the flow. The black line represents the spectrum of the total field, i.e. $E_b=E_{b,s}+ E_{b,a}$ and $E_u=E_{u,s}+ E_{u,a}$, at the time of the maximum dissipation rate.
It is important to note the asymmetric part of the flow is always smaller than the symmetric part for all scales. Thus, the symmetric part of the flow reproduces the $k^{-2}$ energy spectrum, while the asymmetric part of the spectrum evolves passively. 
%At $t=0$ the asymmetric part of the fields has a flat spectrum that
%is contained in the wave numbers [2,10].
%As time progresses the asymmetric part peaks at small scales showing a 
%positive slope with $E_a(k) \sim k^{2}$.  

The results of this section show that no matter how small the amplitude $\rho$ of 
the perturbation added in the TG initial conditions there is a $Re$ for which
the perturbation will grow significantly enough for it to play a (non-linear) dynamical role
in the system. This critical amplitude can be estimated from our
runs to be either $\rho_{crit} = C_1 Re^{-1}$ if energy estimates are considered 
or $\rho_{crit} = C_2 Re^{-3/2}$ if dissipation estimates are considered, where $C_1$ and $C_2$ are constants. % that depend on initial conditions. %The coefficients $c_1,c_2$ depend on the choice of amplitude for which one considers the perturbation as being non-linear.
Consequently, the results explain why symmetry breaking was not observed 
at $t_{peak}$ in the simulations of \cite{leeetal10,pouquetetal10,da13a,da13b} due to the presence of numerical noise. Simulations using single precision accuracy introduce perturbations of amplitude 
$\rho \sim 10^{-8}$ and thus $Re \sim 10^8$ would be required for the symmetries to break in the large scales and $Re \sim 10^5$ for the symmetries to break in the small 
scales. Simulations at such Reynolds numbers cannot be performed on today's largest supercomputers even at single precision accuracy. Therefore, this would make the observation of 
symmetry breaking by numerical noise alone impossible at $t_{peak}$.
We note, however, that symmetry breaking can be observed at later times
as it was shown in \cite{stawarzetal12}.

%%%%%%%%%%%%%%%%%%%%%%%%%%%%%%%%%%%%%%%%%%%%%%%%%%
\section{\label{sec:largepert}Growth of finite amplitude perturbations}
%%%%%%%%%%%%%%%%%%%%%%%%%%%%%%%%%%%%%%%%%%%%%%%%%%

The growth of infinitesimal perturbations gives us a lot of information on the growth of symmetry breaking perturbations. However, it does not provide us with any information about 
a possible change in the spectral exponent and the return or not to a universal behaviour.
For this reason we have performed two series of simulations
with perturbation amplitude $\rho=0.01$ and $\rho=0.1$ (see Table \ref{tbl:dnsparam}). For these values of $\rho$ 
the perturbation grows sufficiently large at $t_{peak}$ to play a dynamical role
in the evolution of the flow.
The wavenumber range of the initial conditions of $\bm u_p$ and $\bm b_p$ was limited
within $2\le | \bm k | \le 4$. The energy ratio $\rho$ and the 
dissipation ratio $\rho_\epsilon$ at $t=0$ are
$\rho_\epsilon \simeq 0.54$ for the $\rho=0.1$ runs and
$\rho_\epsilon \simeq 0.054$ for the $\rho=0.01$ runs.

%%%%%%%%%%%%%%%%%%%%%%%%%%%%%%
\subsection{Temporal behavior}
%%%%%%%%%%%%%%%%%%%%%%%%%%%%%%

\begin{figure}[!ht]
   \begin{subfigure}{8cm}
   \includegraphics[width=\textwidth]{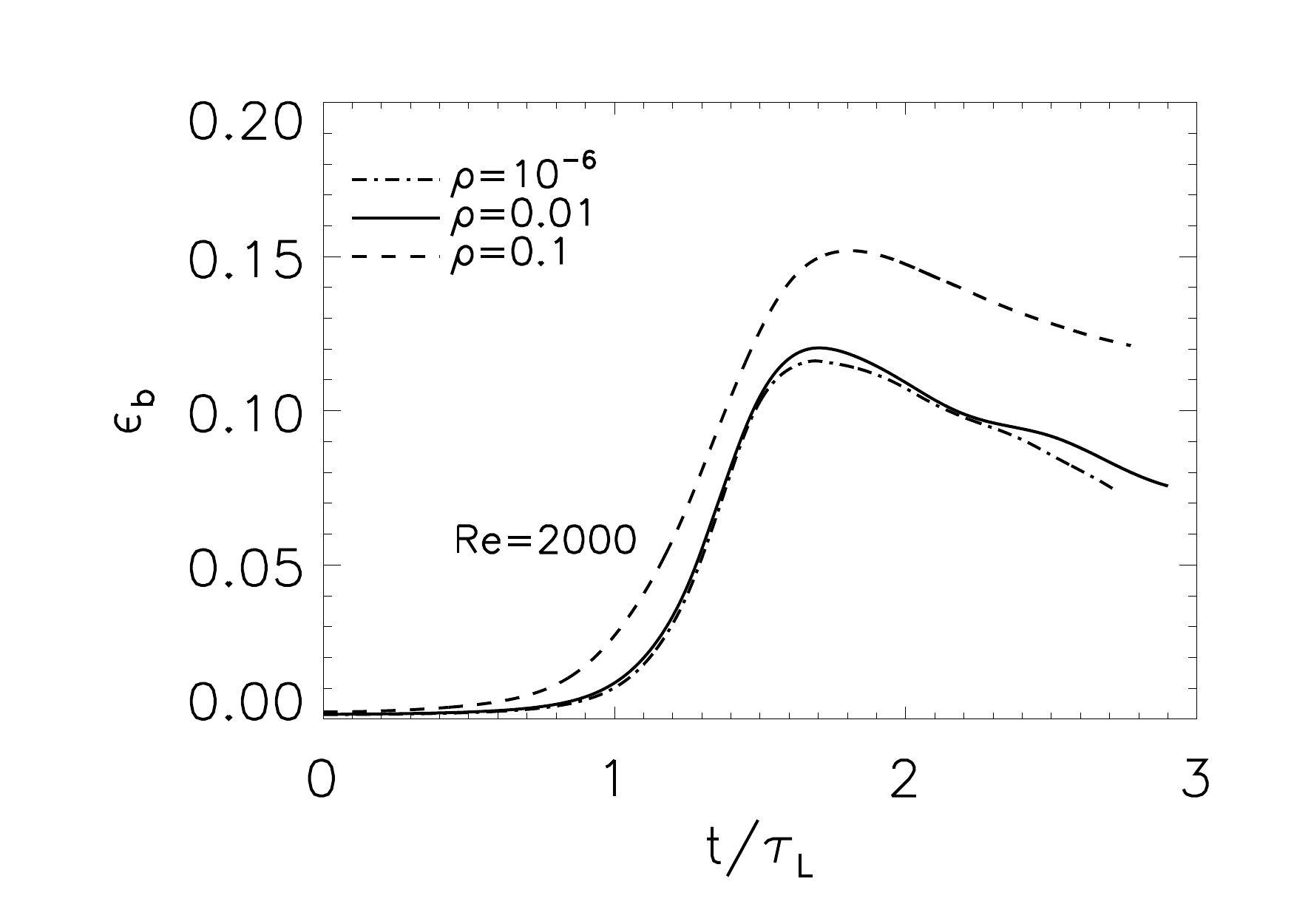}
   \caption{}
  \end{subfigure}
   \begin{subfigure}{8cm}
   \includegraphics[width=\textwidth]{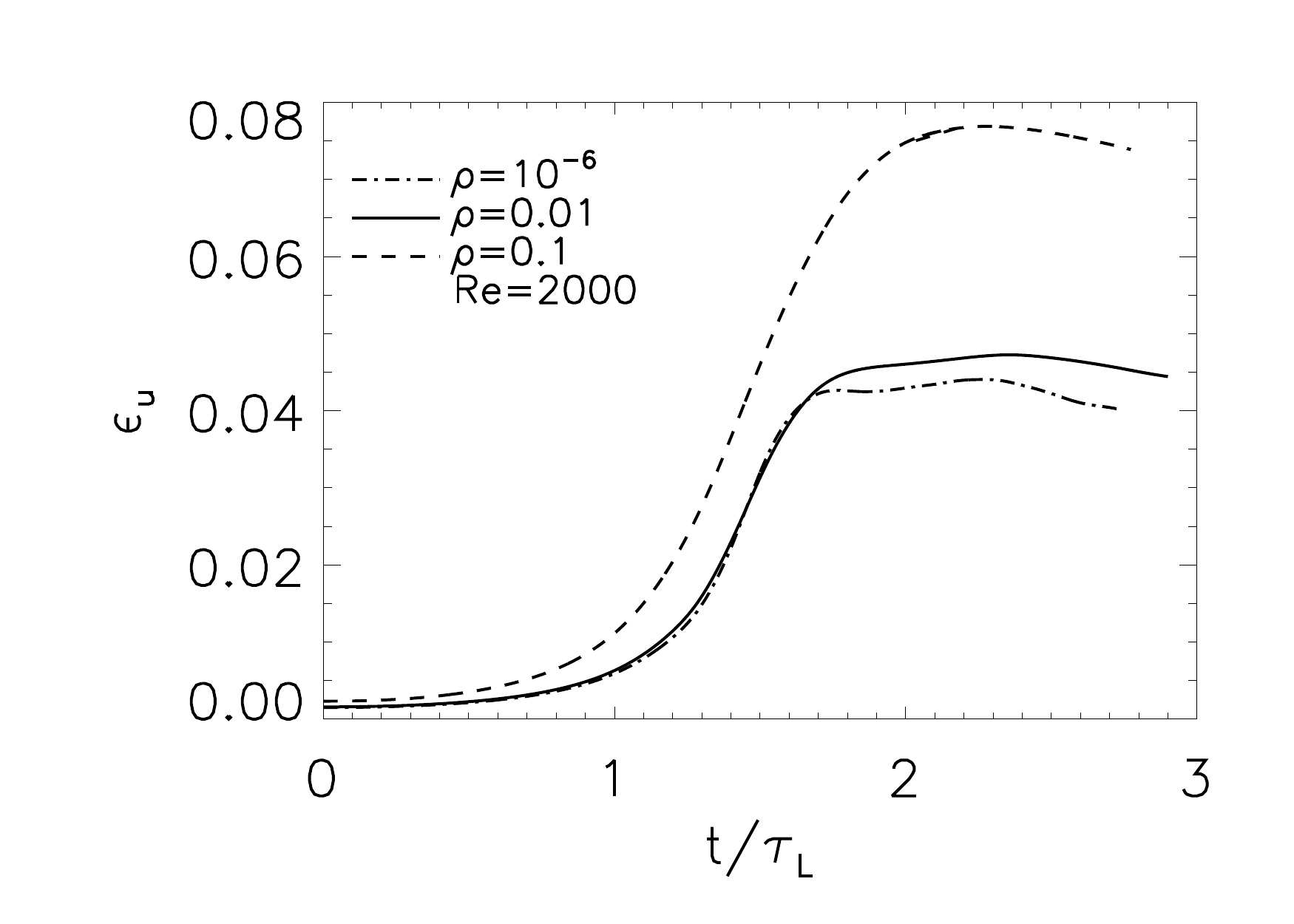}
   \caption{}
  \end{subfigure}
  \caption{(a) Ohmic and (b) viscous dissipation rate as a function of time for $Re=2000$ and three different values of $\rho$.}
  \label{fig:Enst_rho}
 \end{figure}

Figure \ref{fig:Enst_rho}a shows the time evolution of Ohmic dissipation rate $\epsilon_b$ for the runs with $Re=2000$ and three different values of $\rho$. %At the examined Reynolds number the Ohmic dissipation rate forms a peak at $t \simeq 1.7$ for $\rho = 10^{-6}$ and 0.01 cases. 
One can notice that the perturbation of amplitude $\rho=0.1$ has %already 
significantly changed the time evolution of $\epsilon_b$, which has increased in amplitude and its peak has been shifted later in time (i.e. $t_{peak}/\tau_L \simeq 1.8$). Smaller variations are observed for the $\rho=0.01$ case with a small increase of $\epsilon_b$ ($\sim 5\%$) and no visible change in $t_{peak}/\tau_L \simeq 1.7$ in comparison to the $\rho = 10^{-6}$ case.
Moreover, a new local peak starts to form around $t/\tau_L \simeq 2.4$ that is not present for the $\rho=10^{-6}$ run. This new peak, as we will show later, is due to the formation of smaller scales by the breaking of the symmetries.

In Fig. \ref{fig:Enst_rho}b we observe that the viscous dissipation rate $\epsilon_u$ peaks at later times than $\epsilon_b$ at this $Re$. For the run with perturbation amplitude $\rho=0.1$ the peak of the viscous dissipation rate has increased by $40\%$ while for $\rho=0.01$ there is a $10\%$ increase with reference to the $\rho=10^{-6}$ case. It is worth noting that the peak of $\epsilon_u$ for the $\rho=0.01$ case coincides with the second local peak of $\epsilon_b$ that takes place at $t/\tau_L \simeq 2.4$.

\begin{figure}[!ht]
    \begin{subfigure}{8cm}
   \includegraphics[width=\textwidth]{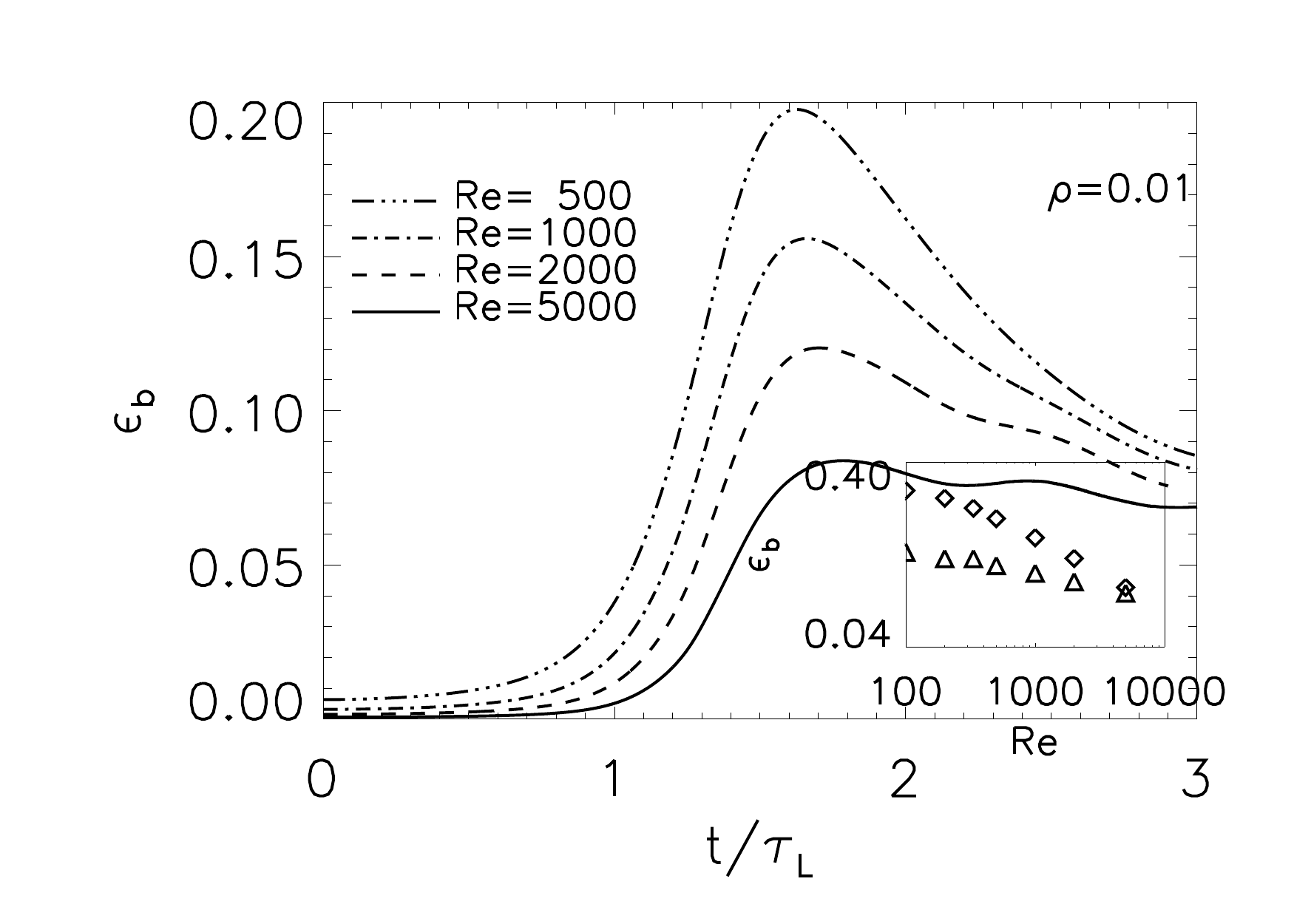}
   \caption{}
  \end{subfigure}
   \begin{subfigure}{8cm}
   \includegraphics[width=\textwidth]{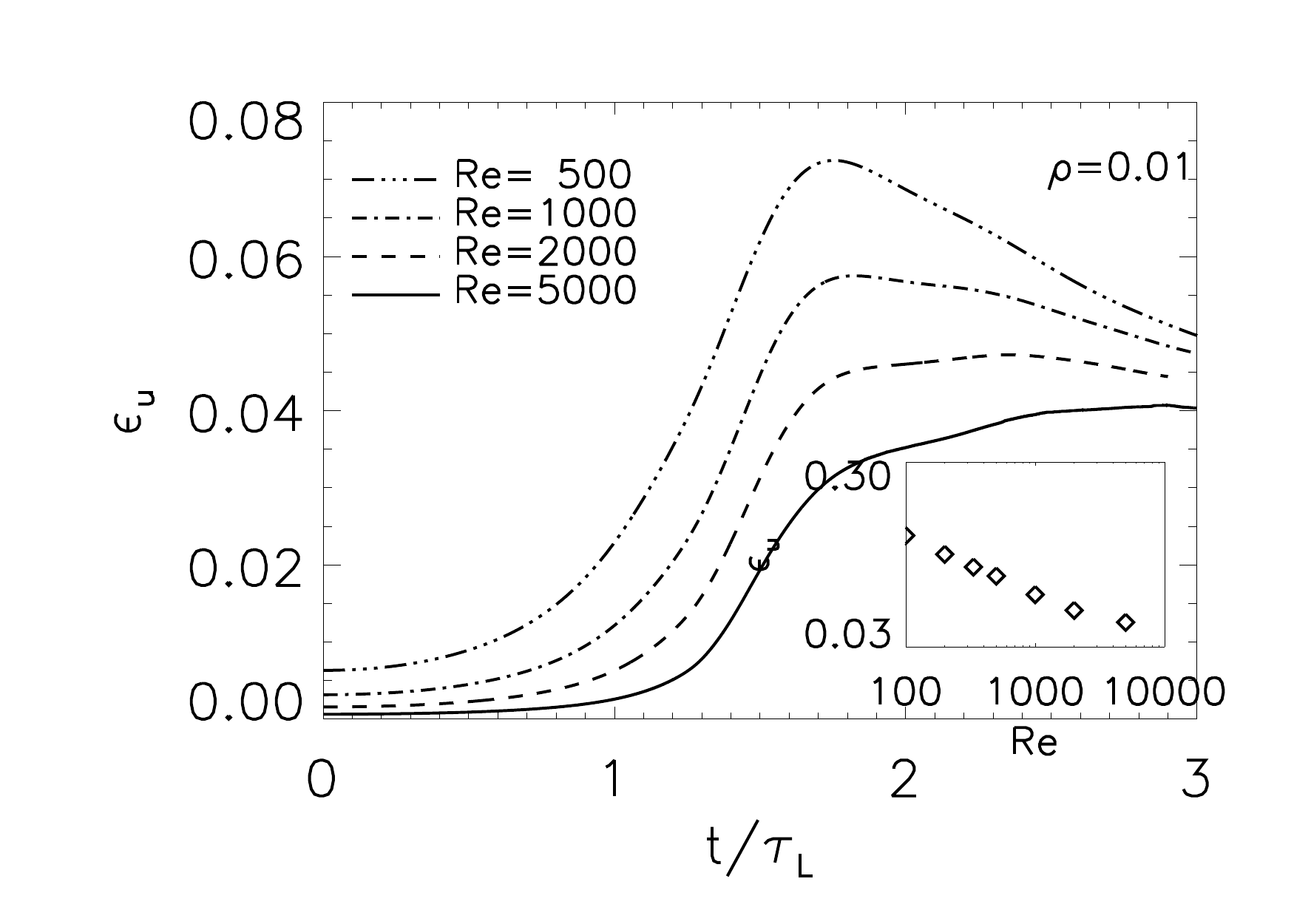}
   \caption{}
  \end{subfigure}
  \caption{(a) Ohmic and (b) viscous dissipation rate as a function of time for $\rho=0.01$ and different values of $Re$.}
  \label{fig:Enst_Re}
 \end{figure}

The occurrence of the new local peak of Ohmic dissipation can be seen more clearly at higher Reynolds numbers. Figure \ref{fig:Enst_Re}a presents the dissipation rates as a function of time for four different $Re$ and $\rho=0.01$.
For small values of the Reynolds number a single peak appears for the time evolution of $\epsilon_b$ at $t/\tau_L \simeq 1.7$ that coincides with the time of the dissipation peak observed in the unperturbed system. As the Reynolds number is increased a new peak appears at $t/\tau_L \simeq 2.4$. The second peak can be seen clearly only for the $Re = 5000$ run and it is very close in amplitude with the first peak at $t/\tau_L \simeq 1.7$. Hence, the role played by this small perturbation is only apparent at very large $Re$, while its effect is muffled at smaller $Re$.
On the other hand, a single peak is developed in the evolution of the viscous dissipation rate $\epsilon_u$, which occurs at $t/\tau_L \simeq 1.8$ (a little later than the Ohmic dissipation peak)
for $Re \le 1000$ (see Fig. \ref{fig:Enst_Re}b) but moves further in time at $t/\tau_L \simeq 2.4$ and 2.9 for $Re = 2000$ and 5000, respectively.

%The insets in Figs. \ref{fig:Enst_Re}a and \ref{fig:Enst_Re}b illustrate the values of the dissipation rates for different Reynolds numbers. 
The inset in Fig. \ref{fig:Enst_Re}a illustrates the values of $\epsilon_b$ at the moment of the first peak $t/\tau_L = 1.7$ (diamonds) and of the second peak at $t/\tau_L = 2.4$ (triangles) for different Reynolds numbers. The second peak of $\epsilon_b$ seems to reach each asymptotic state much faster than the first peak as $Re$ increases with $Re = 5000$ the transitional point where $\epsilon_b|_{t/\tau_L=1.7} \simeq \epsilon_b|_{t/\tau_L=2.4}$. Then, for $Re \gg 1$ the trends of the two peaks of $\epsilon_b$ suggest that the second peak will become dominant and Reynolds number independent. The maximum values of $\epsilon_u$ that are plotted in the inset of Fig. \ref{fig:Enst_Re}b as a function of $Re$ also indicate that the viscous dissipation rate is far from reaching its asymptotic state even for our highest resolution simulations ($Re = 5000$), which are at the limit of the current computational power. It is interesting that the Ohmic and viscous dissipation obey different 
high Reynolds number 
asymptotics. In other words, the small scales of the magnetic field seem to reach its universal regime at lower $Re$ than the small scales of the velocity field. % and this might be an indication of non-universality...?

%%%%%%%%%%%%%%%%%%%%%%%%%%%%%%%%%%%%%%%%%%%%%
\subsection{Symmetry breaking and structures}
\begin{figure}[!ht]
  \begin{subfigure}{8cm}
   \includegraphics[width=\textwidth]{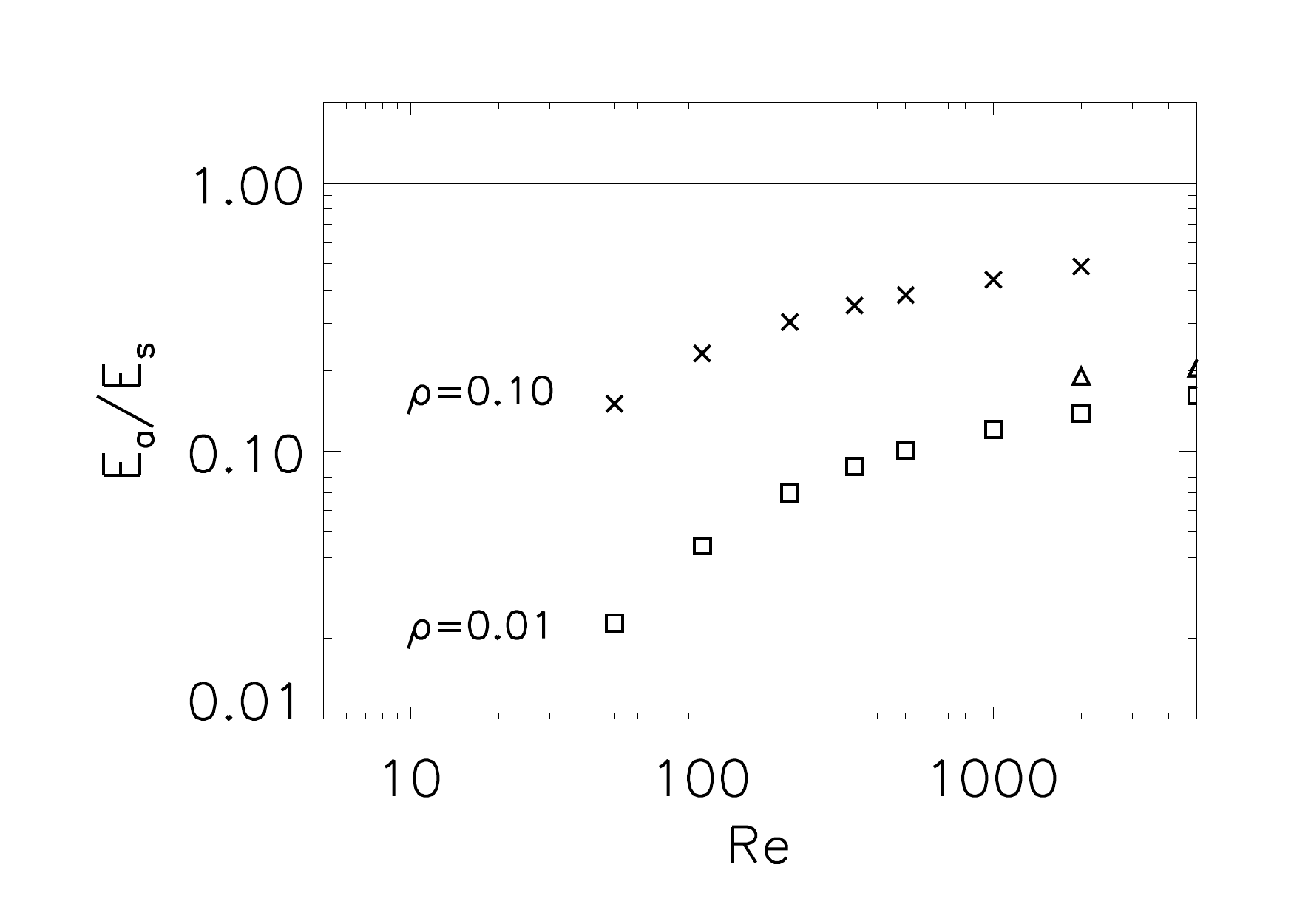}
   \caption{}
  \end{subfigure}
  \begin{subfigure}{8cm}
   \includegraphics[width=\textwidth]{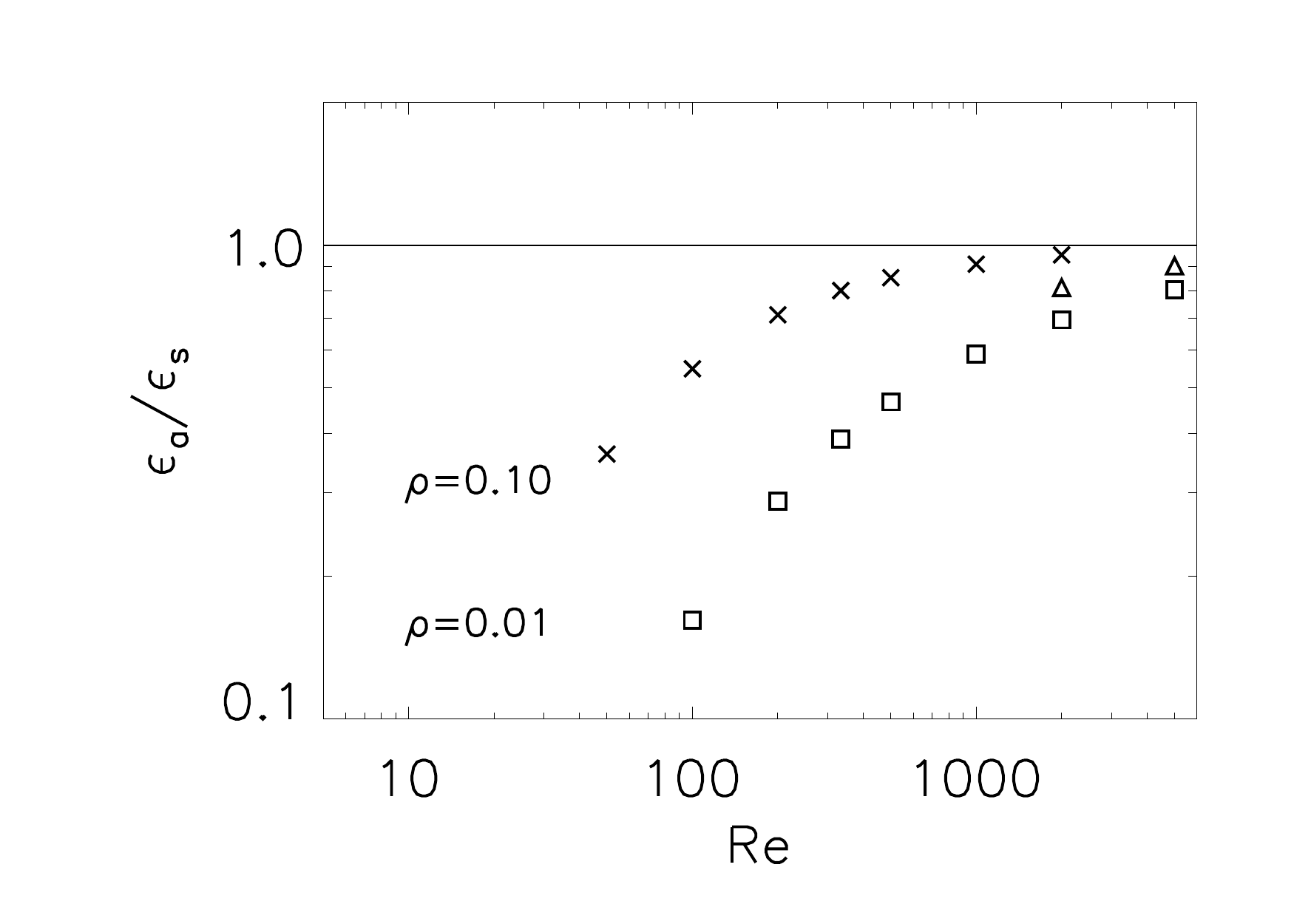}
   \caption{}
  \end{subfigure}
  \caption{(a) Energy ratio $E_a/E_s$ and (b) dissipation ratio $\epsilon_a/\epsilon_s$
           as functions of $Re$ at the time of maximum dissipation rate for $\rho=0.01$ and 0.1.
           The triangles correspond to the run with $\rho=0.01$ for the second peak of dissipation
           that only appeared at the high $Re$ cases at $t/\tau_L \simeq 2.4$.}
  \label{fig:Ener_rat_nl}
 \end{figure}

%%%%%%%%%%%%%%%%%%%%%%%%%%%%%%%%%%%%%%%%%%%%%
%
In Figs. \ref{fig:Ener_rat_nl}a and \ref{fig:Ener_rat_nl}b we show the values of the energy ratio  $E_a/E_s$ and the dissipation ratio $\epsilon_a/\epsilon_s$, respectively, at the peak of the energy dissipation rate for all the different $Re$ that we consider in this study for runs with $\rho=0.01$ and 0.1 (see Table \ref{tbl:dnsparam}).
The effect of non-linearity is evident,
since both cases deviate from the scalings $E_a/E_s \sim Re$ and $\epsilon_a/\epsilon_s \sim Re^{3/2}$ observed in section \ref{sec:smallpert}.
In particular, at high $Re$ the asymmetric part of the energy for the runs with 
$\rho=0.1$ appears to asymptote towards $E_a \simeq 0.5 E_s$ while in the $\rho=0.01$
it is significantly smaller, i.e. $E_a \simeq 0.15 E_s$,
with slightly higher value at the second dissipation peak (see Fig. \ref{fig:Ener_rat_nl}a). 
This implies that at the large, energy containing scales only a 
modest breaking of the symmetries has occurred for $\rho=0.01$ runs.
In the small scales, however, the symmetries seem to be fully broken
for both values of $\rho$. For the $\rho=0.1$ case
the dissipation ratio reaches values close to unity even for $Re=500$ and for $\rho=0.01$ 
we have $\epsilon_a \simeq 0.8 \epsilon_s$ at the first dissipation peak and 
              $\epsilon_a \simeq 0.9 \epsilon_s$ at the second dissipation peak for $Re=5000$.

Symmetry breaking of the resulting structures can be also realised through visualisations. While three dimensional images of the full computational box provide global information, they are sensitive in the choice of iso-contour levels and very often can be misleading. We have thus chosen to show colour plots of two dimensional slices that pass through the high current and vorticity density regions.

%%%%%%%%%%%%%%%%%%%%%%%%%%%%%%%%%%%%%%%%%%%%%%%%%%%
\begin{sidewaysfigure*}
\vspace{11.3cm}
\begin{subfigure}{0.245\linewidth}
\includegraphics[height=5.7cm]{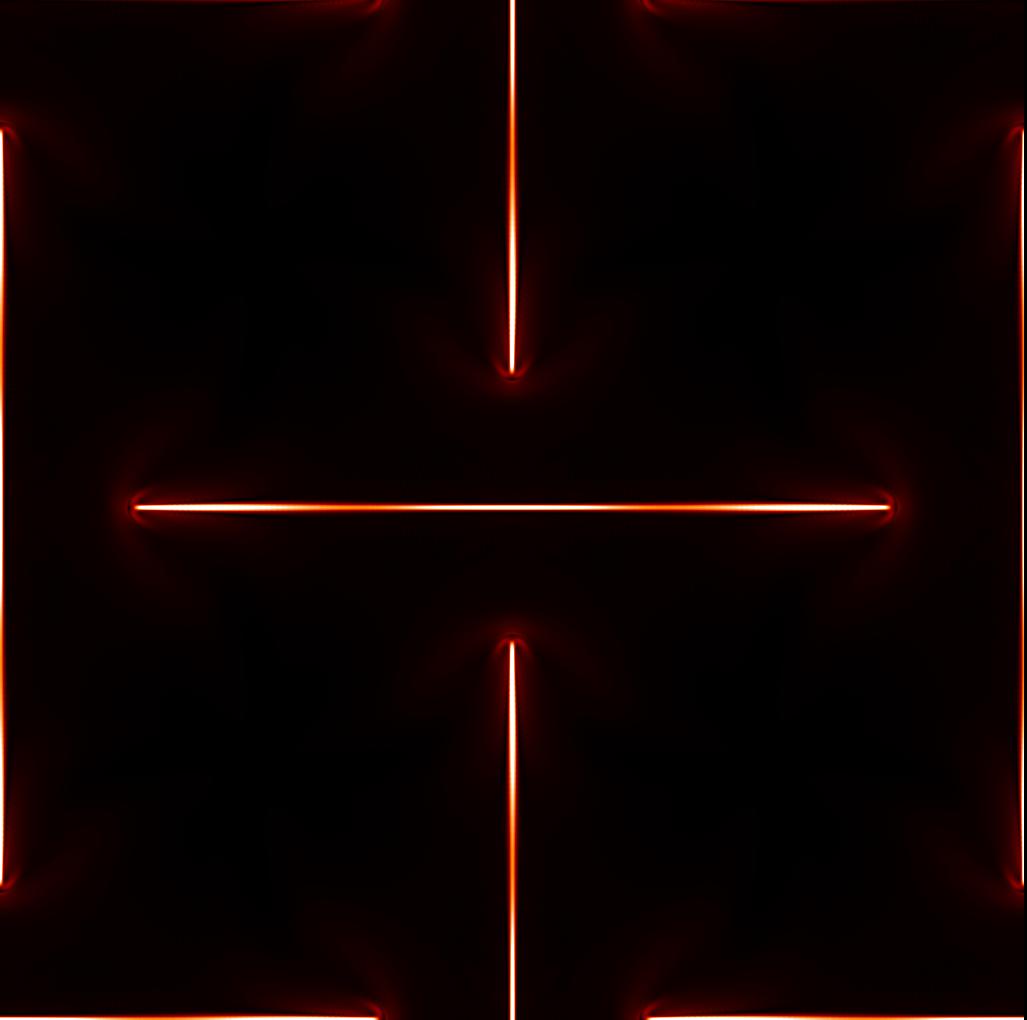}
\end{subfigure}
\begin{subfigure}{0.245\linewidth}
\includegraphics[height=5.7cm]{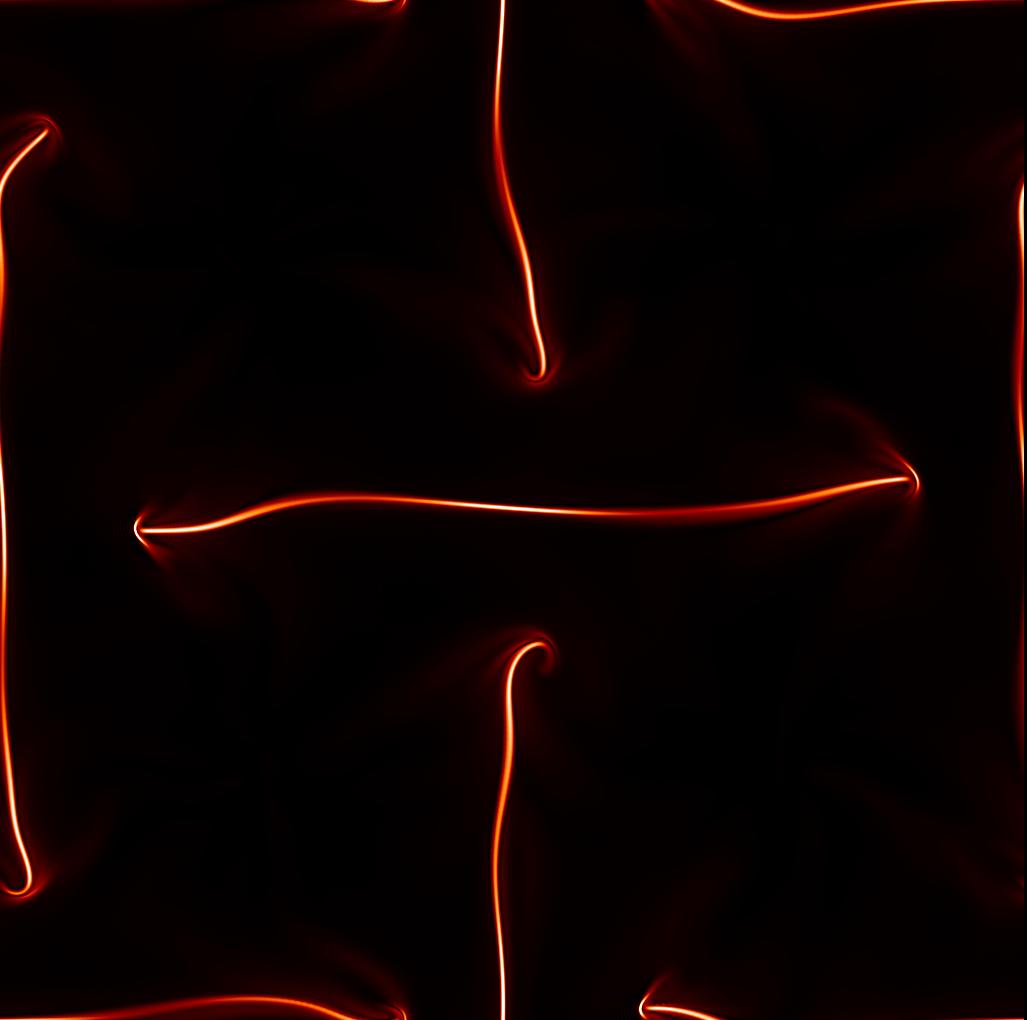}
\end{subfigure}
\begin{subfigure}{0.245\linewidth}
\includegraphics[height=5.7cm]{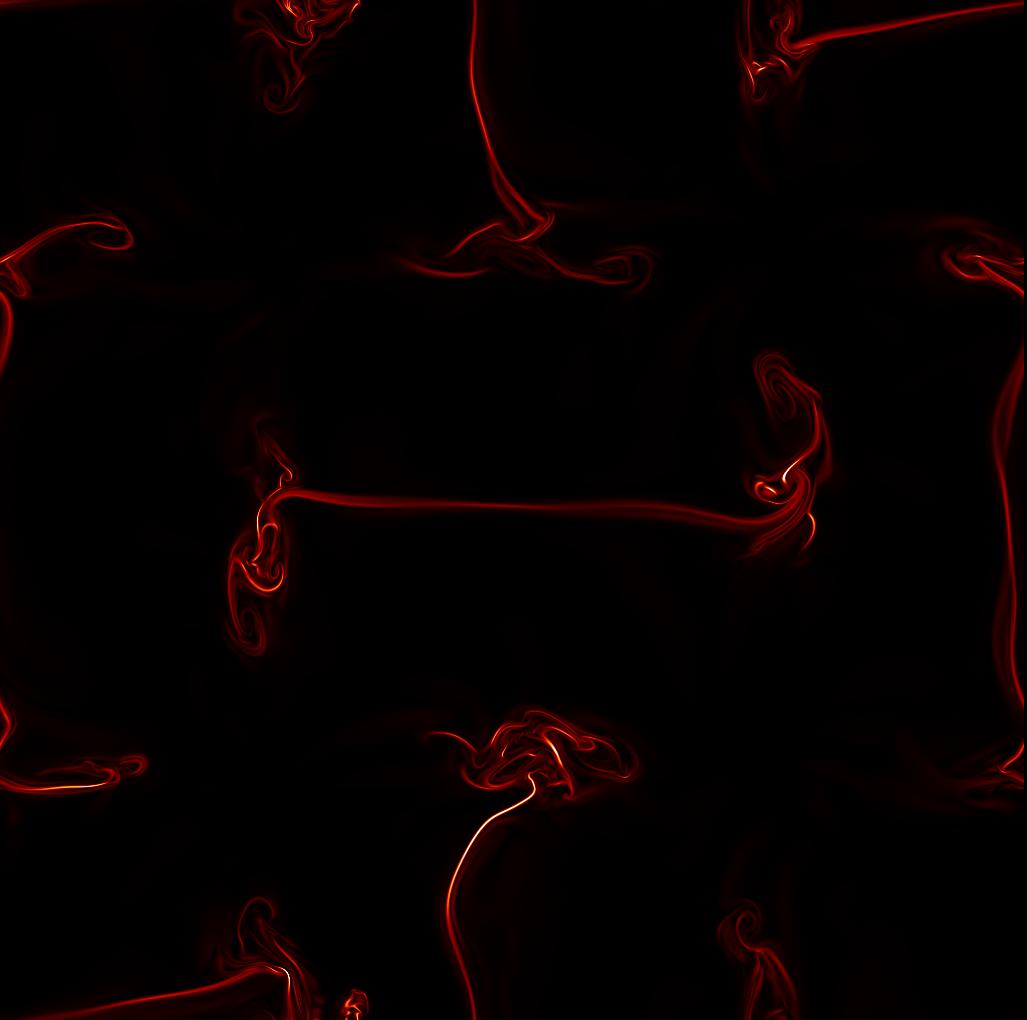}
\end{subfigure}
\begin{subfigure}{0.245\linewidth}
\includegraphics[height=5.7cm]{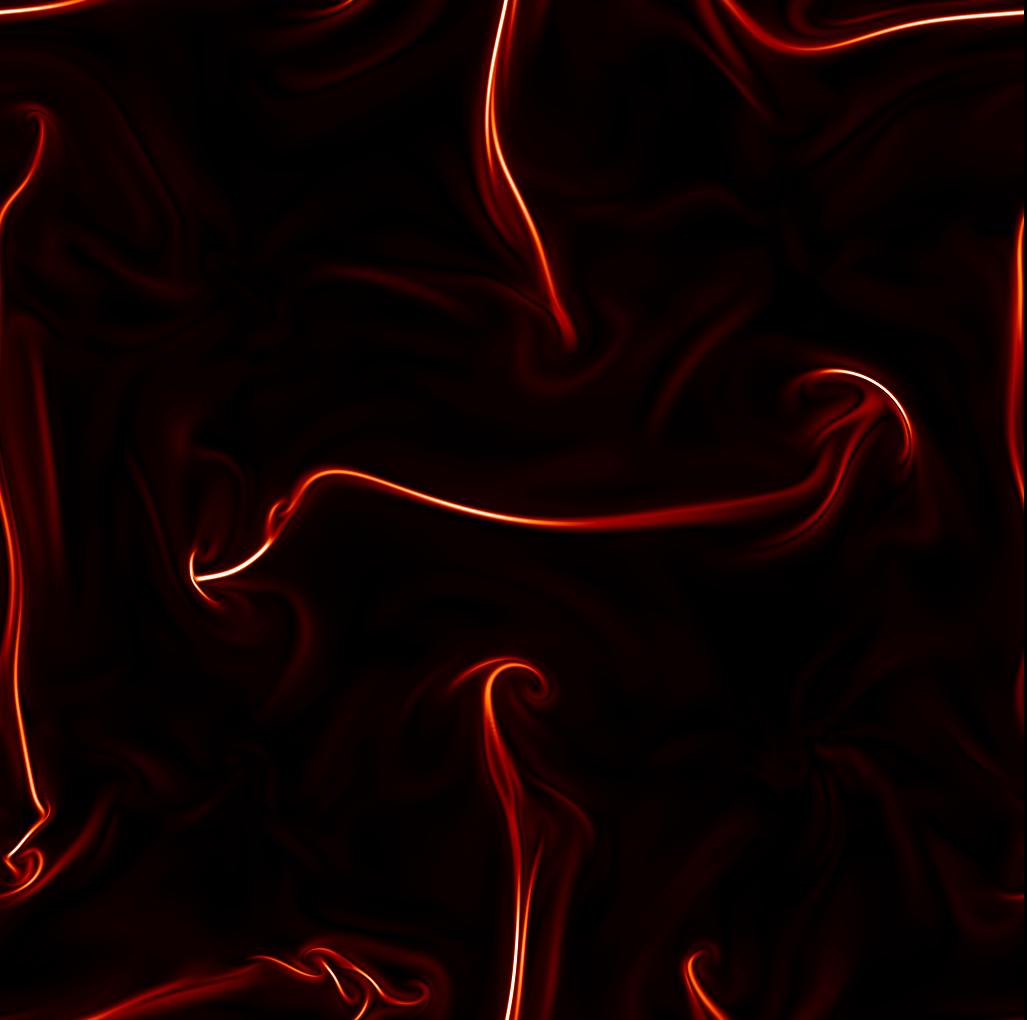}
\end{subfigure}

\vspace{0.7cm}

\begin{subfigure}{0.245\linewidth}
\includegraphics[height=5.7cm]{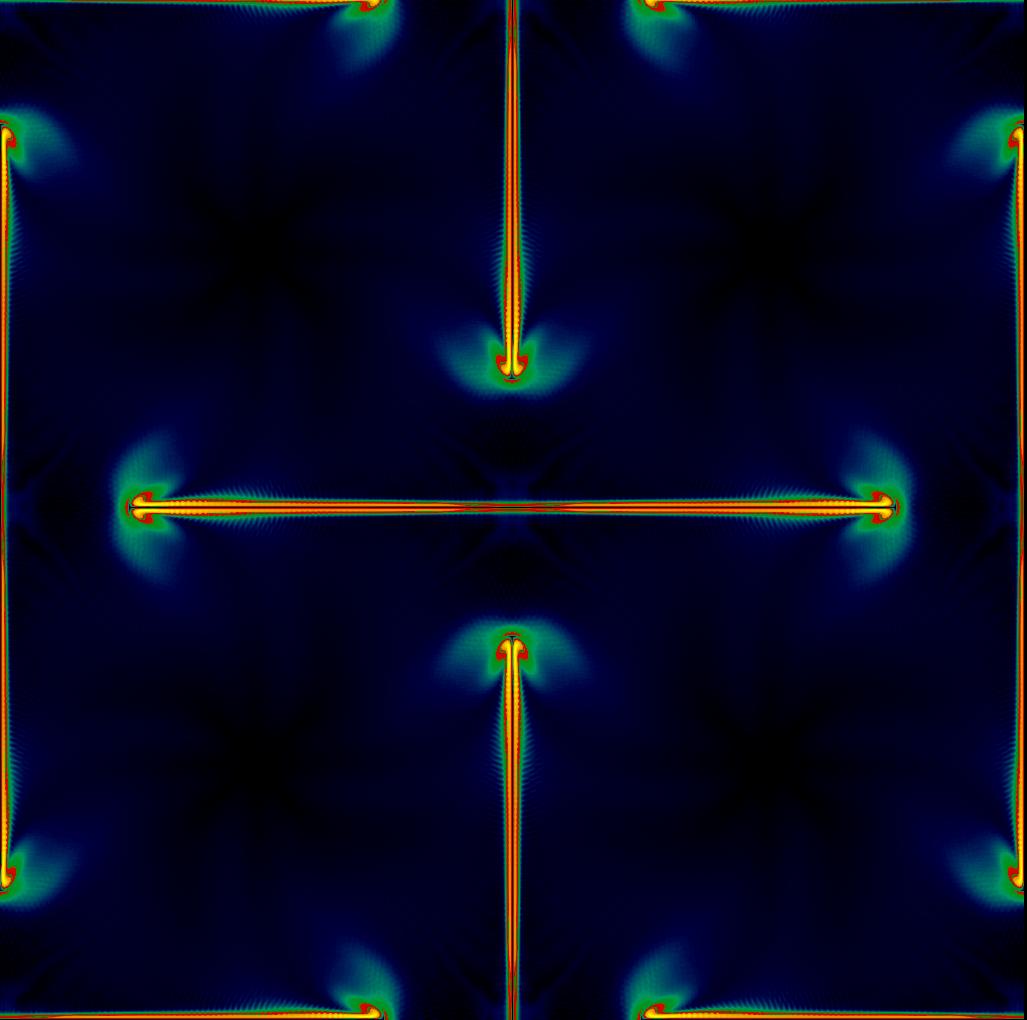}
\caption{}
\end{subfigure}
\begin{subfigure}{0.245\linewidth}
\includegraphics[height=5.7cm]{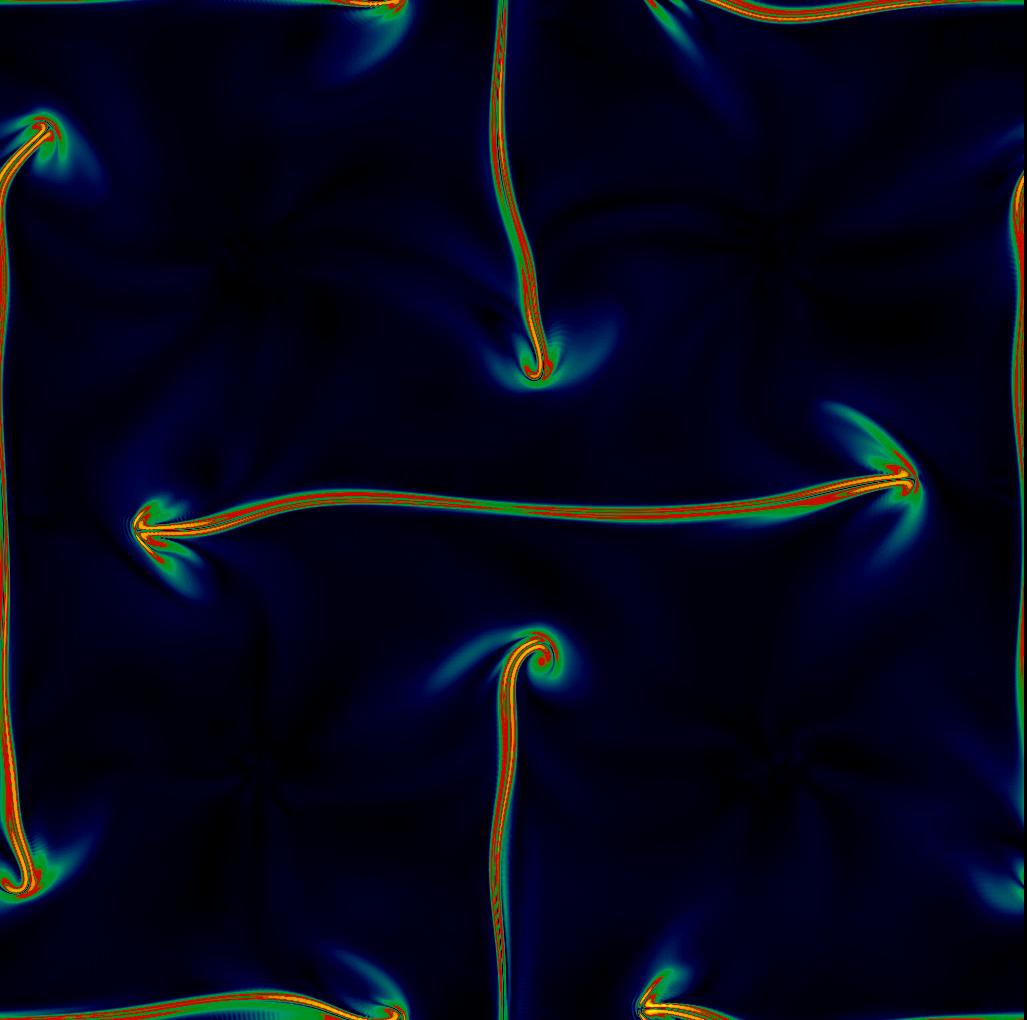}
\caption{}
\end{subfigure}
\begin{subfigure}{0.245\linewidth}
\includegraphics[height=5.7cm]{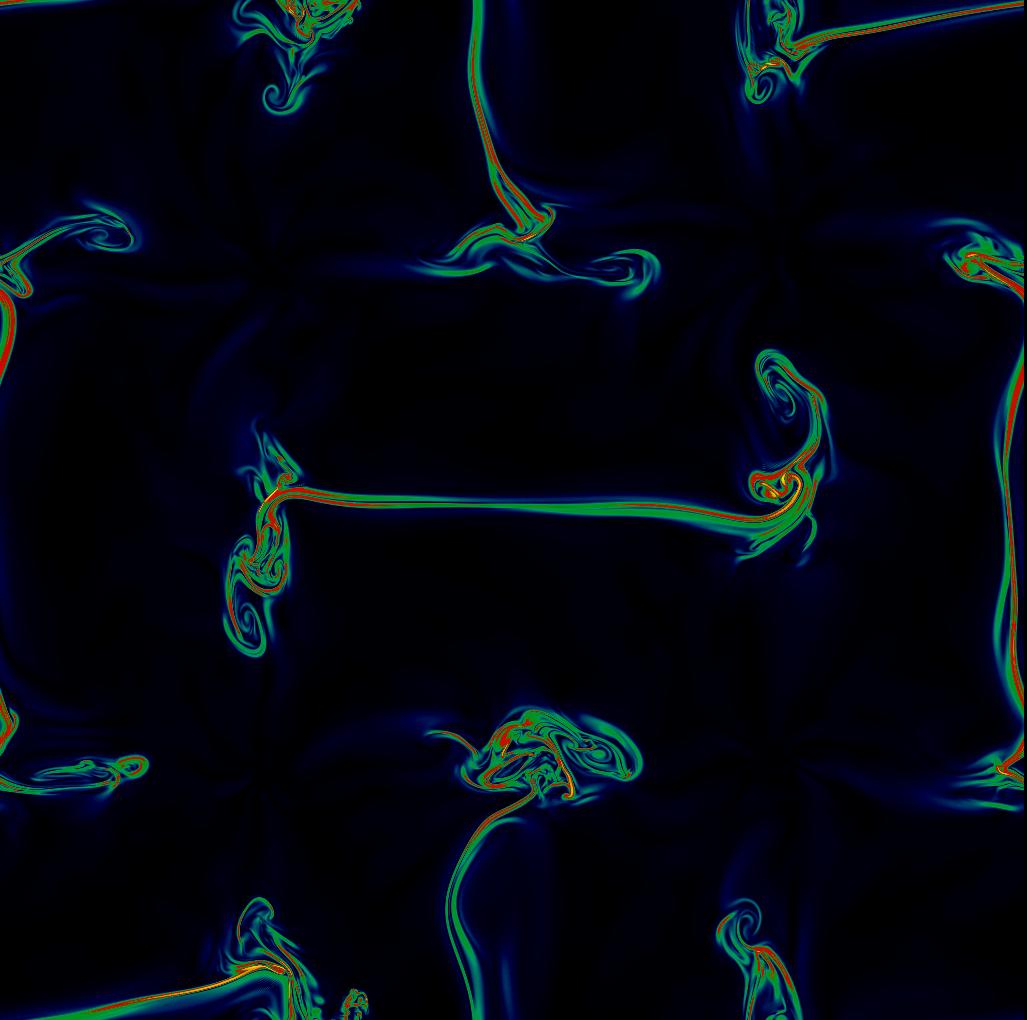}
\caption{}
\end{subfigure}
\begin{subfigure}{0.245\linewidth}
\includegraphics[height=5.7cm]{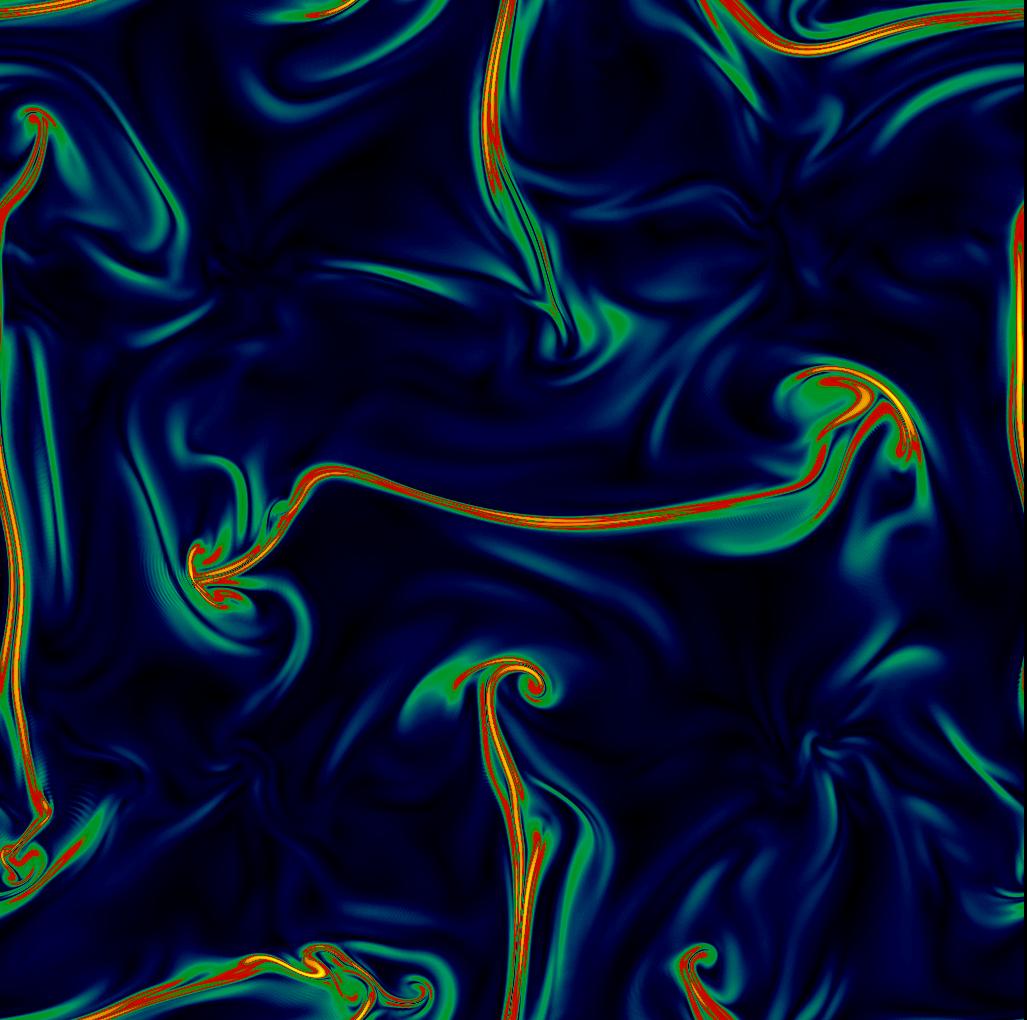}
\caption{}
\end{subfigure}
\caption{(Color online) Current (top panels) and vorticity (bottom panels) density at the time of maximum dissipation rate for
           (a) $\rho=10^{-6}$, $Re=2000$,
           (b) $\rho=0.01$, $Re=2000$,
           (c) $\rho=0.01$, $Re=5000$ and
           (d) $\rho=0.1$, $Re=2000$.}
\label{fig:colorfigs}
\end{sidewaysfigure*}
%%%%%%%%%%%%%%%%%%%%%%%%%%%%%%%%%%%%%%%%%%%%%%%%%%%

Figure \ref{fig:colorfigs} illustrates the current (top panels) and vorticity (bottom panels) density at the $z = \pi/4$ plane. In Fig. \ref{fig:colorfigs}a we show results for the $\rho=10^{-6}$ and $Re=2000$ run where the perturbation evolved passively. In this case, the strong current and vorticity sheets that appear at the $x=0,\pi/2$ and $y=0,\pi/2$ planes
are responsible for the $k^{-2}$ energy spectrum (see \cite{da13b}). Therefore, the stability of these structures is crucial to determine the presence or absence of universality. The effect of the perturbation on the structures becomes pronounced for the $\rho=10^{-2}$ and $Re=2000$ case (see Fig. \ref{fig:colorfigs}b). The development of the perturbation has lead to the bending and curling of the current and vorticity sheets. The basic structures, however, remain unaltered without development of additional features. On the other hand, at higher Reynolds number (i.e. $Re=5000$ and $\rho=10^{-2}$) more structures appear (see Fig. \ref{fig:colorfigs}c). While the bended current and vorticity sheets are still present,
the flow at their edge has been fully ``randomised" by the development of the instability. At this location, structures with no particular order and reminiscent of flows with random initial conditions have formed. Note that these ``random" structures have generated scales smaller than the thickness of the ``ordered" current/vorticity sheets. These structures are possibly responsible for the second peak of the Ohmic dissipation observed in Fig. \ref{fig:Enst_Re}a but also for the peak of the viscous dissipation at later times as $Re$ increases (see Fig. \ref{fig:Enst_Re}b).
This observation supports our conjecture that at $Re \gg 1$ these turbulent fluctuations will dominate and the second peak of $\epsilon_b$ will become a global maximum. Finally, in Fig. \ref{fig:colorfigs}d the effect of the perturbation is much more pronounced for the $\rho=0.1$ and $Re=2000$ run; not only at the edge of the current and vorticity sheets but also away from the symmetry planes, where strong current and vorticity turbulent structures have emerged. This probably indicates that the amplitude of the perturbation $\rho=0.1$ was large enough that asymmetric currert/vorticity was not only amplified by its interaction with the symmetric part of the flow but also by self-interaction of the structures introduced by the perturbation.

%%%%%%%%%%%%%%%%%%%%%%%%%%%%%%%%%%%%%%%%%%%%%
%\subsection{Energy spectra}
\subsection{Spectral behaviour}
%%%%%%%%%%%%%%%%%%%%%%%%%%%%%%%%%%%%%%%%%%%%%

As we stated in the introduction, it is an open question whether the breaking of the symmetries 
will change the power law exponent of the spectra. Therefore, in this section we investigate the effect of symmetry breaking on the inertial range scaling of our energy spectra.

\begin{figure}[!ht]
   \begin{subfigure}{8cm}
   \includegraphics[width=\textwidth]{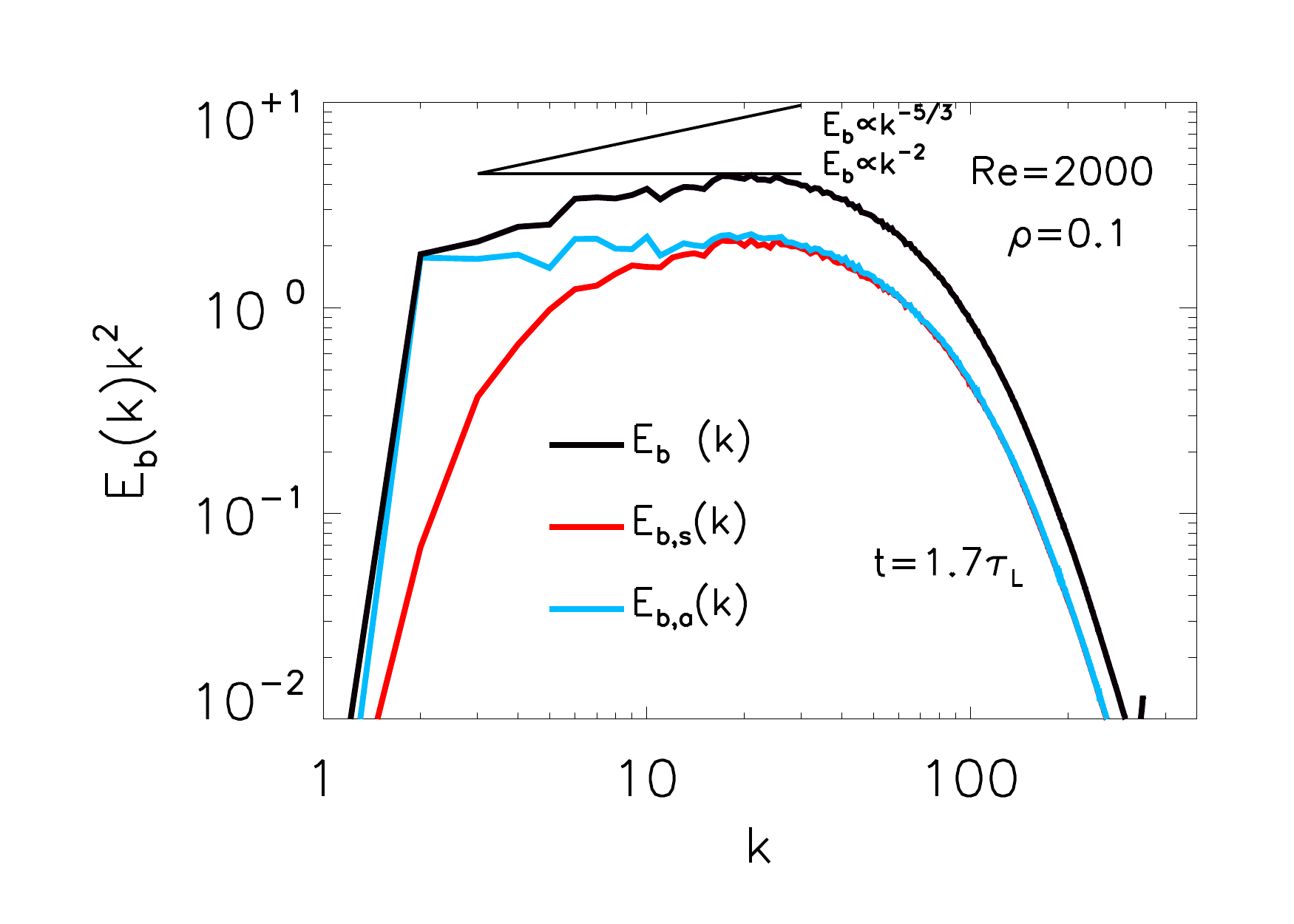}
   \caption{}
  \end{subfigure}
  \begin{subfigure}{8cm}
   \includegraphics[width=\textwidth]{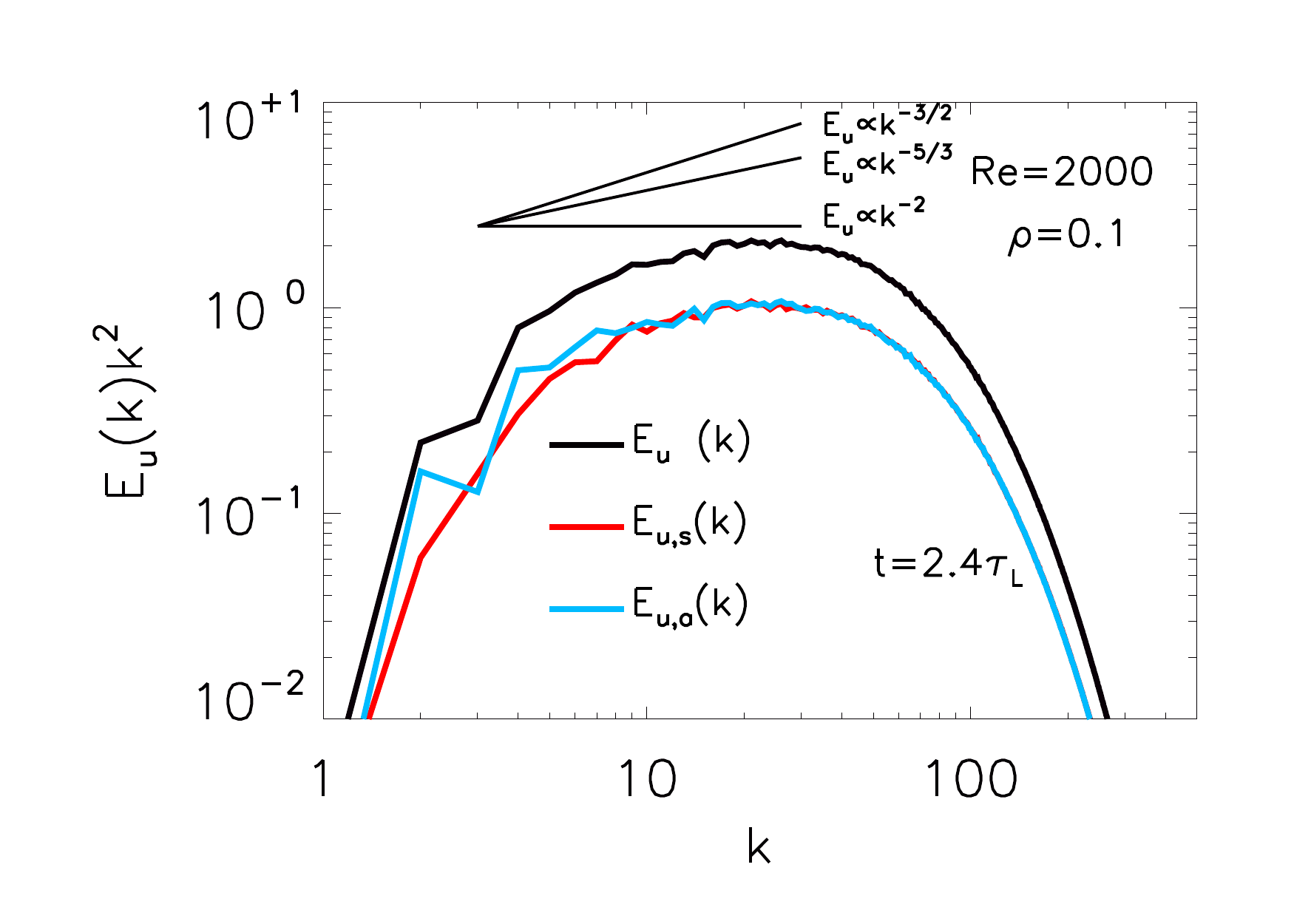}
   \caption{}
  \end{subfigure}
  \caption{(Color online) (a) Magnetic energy and (b) kinetic energy spectra at the peak of Ohmic and viscous dissipation rate, respectively, for $\rho=0.1$ and $Re=2000$.}
  \label{fig:SpecE1}
 \end{figure}

In Fig. \ref{fig:SpecE1} we show the energy spectra for the $\rho=0.1$ and $Re=2000$ case. 
Figure \ref{fig:SpecE1}a represents the magnetic energy spectrum at the peak of the Ohmic dissipation and Fig. \ref{fig:SpecE1}b the kinetic energy spectrum at the peak of viscous dissipation. The blue (light-grey) and the red (dark-grey) lines in these figures indicate the symmetric and asymmetric parts of the energy spectra, respectively, whereas the black lines indicate the full spectra, i.e. symmetric plus asymmetric part. 
%What can be observed is that at short
%times while the symmetric part dominates in the large scales the asymmetric random part dominates 
%in the small scales. As they evolve in time 
At the peak of Ohmic dissipation $E_{b,s}$ and $E_{b,a}$ reach equipartition within the range $20 \lesssim k \leq k_{max}$. The full magnetic energy spectrum compensated by $k^2$ has clearly 
a positive slope and a linear fit indicates a value close to the $k^{-5/3}$ scaling. 
For the kinetic energy spectrum, equipartition occurs between $E_{u,s}$ and $E_{u,a}$ at all
scales. The slope of the full compensated spectrum $k^2 E_u$ is positive and also close to $k^{-5/3}$. However, the range of wave numbers that exhibit a power law for the velocity field is shorter than for the magnetic field and a distinction between $k^{-5/3}$ and $k^{-3/2}$ is not possible.
Hence, it is clear that for $\rho=0.1$ the spectrum moved away from the $k^{-2}$ scaling and it returned to the classical $k^{-5/3}$ (or $k^{-3/2}$) turbulence scaling. Note, however, that this case was strongly perturbed at $t=0$ with a significant amount of enstrophy and current density introduced by the perturbation.

\begin{figure}[!ht]
  \begin{subfigure}{8cm}
   \includegraphics[width=\textwidth]{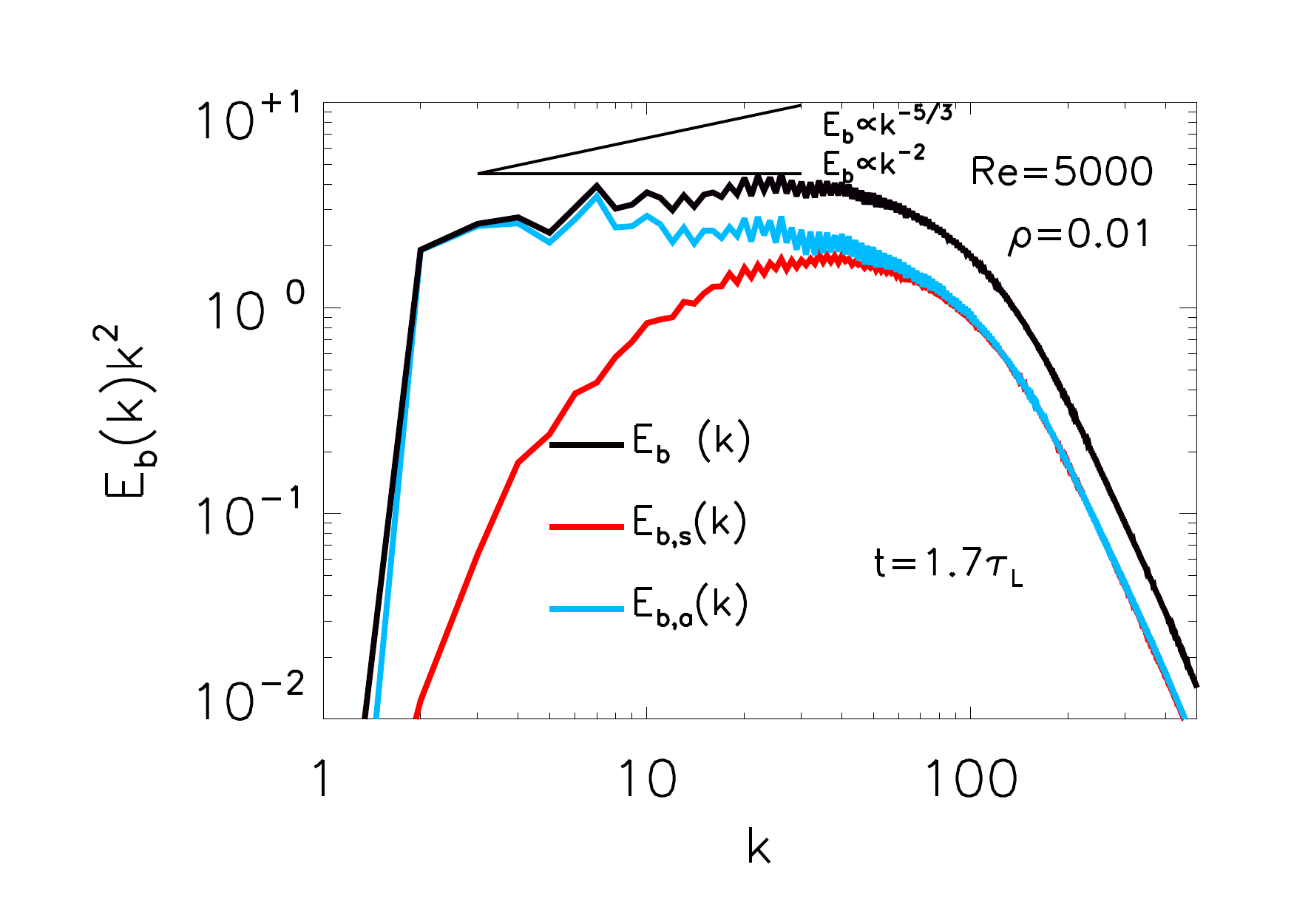}
   \caption{}
  \end{subfigure}
  \begin{subfigure}{8cm}
   \includegraphics[width=\textwidth]{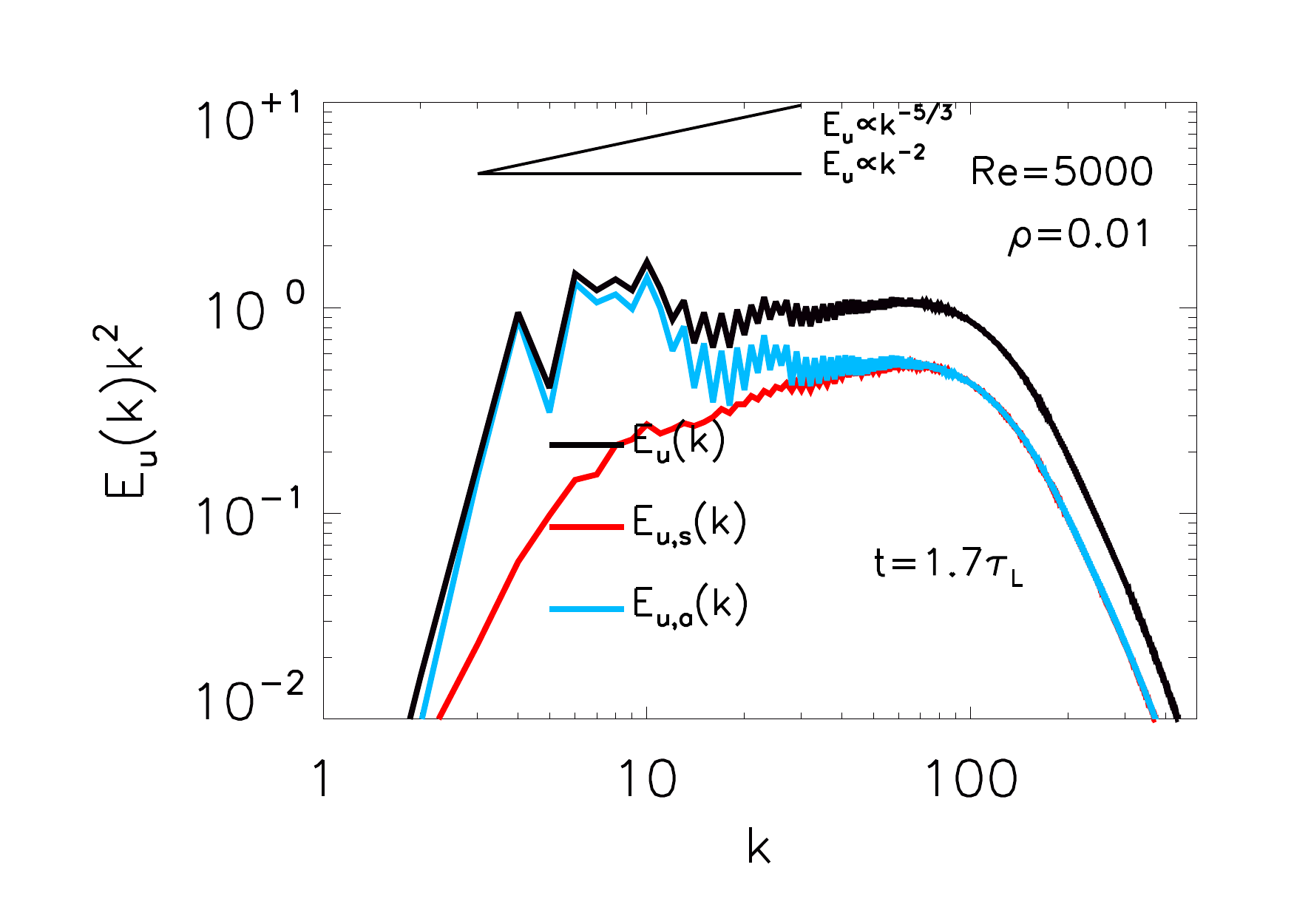}
   \caption{}
  \end{subfigure}
  \caption{(Color online) (a) Magnetic energy and (b) kinetic energy spectra for $\rho=0.01$ and $Re=5000$ at the first peak of total dissipation rate $t/\tau_L=1.7$.}
  \label{fig:SpecE2a}
 \end{figure}

\begin{figure}[!ht]
  \begin{subfigure}{8cm}
   \includegraphics[width=\textwidth]{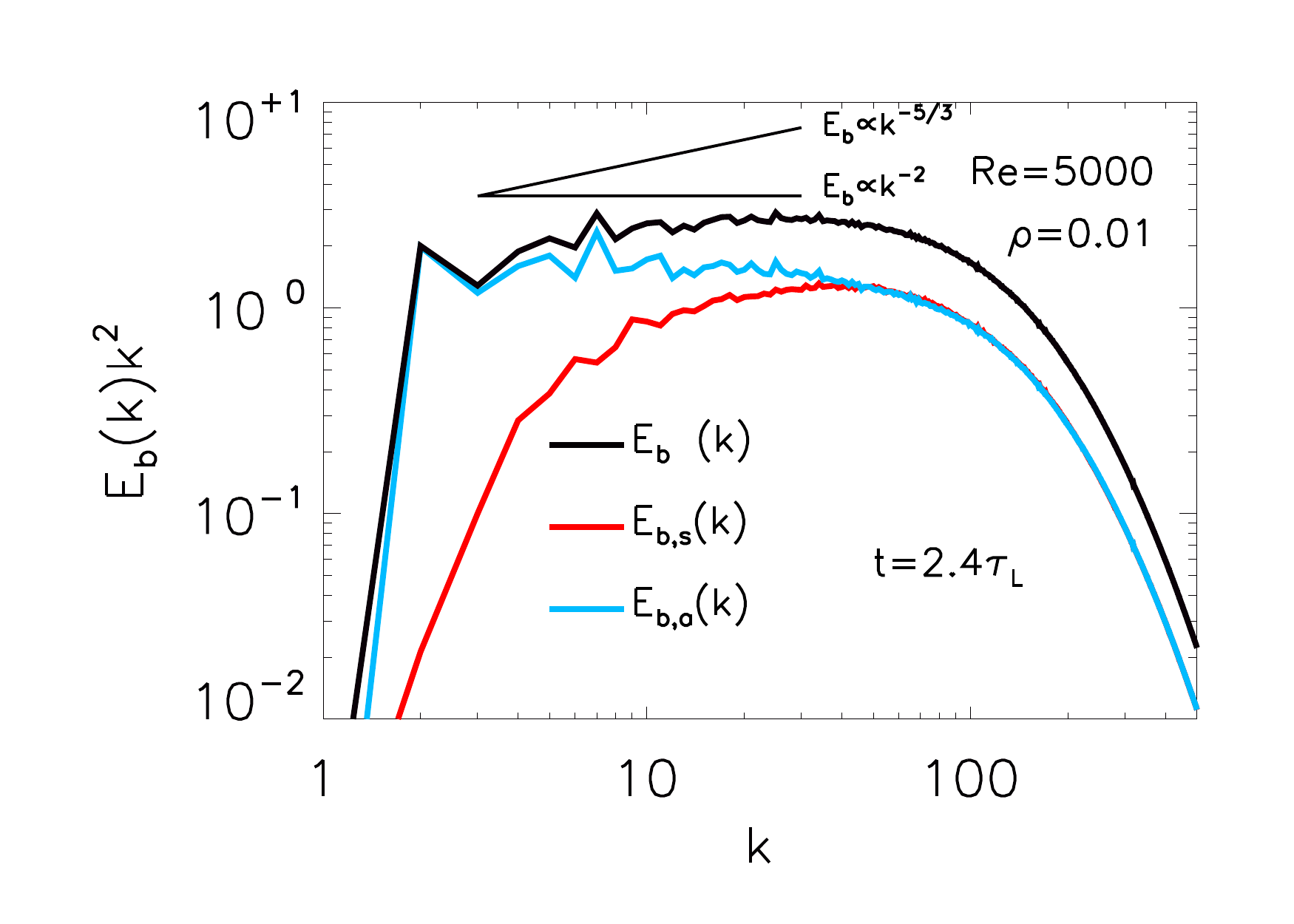}
   \caption{}
  \end{subfigure}
  \begin{subfigure}{8cm}
   \includegraphics[width=\textwidth]{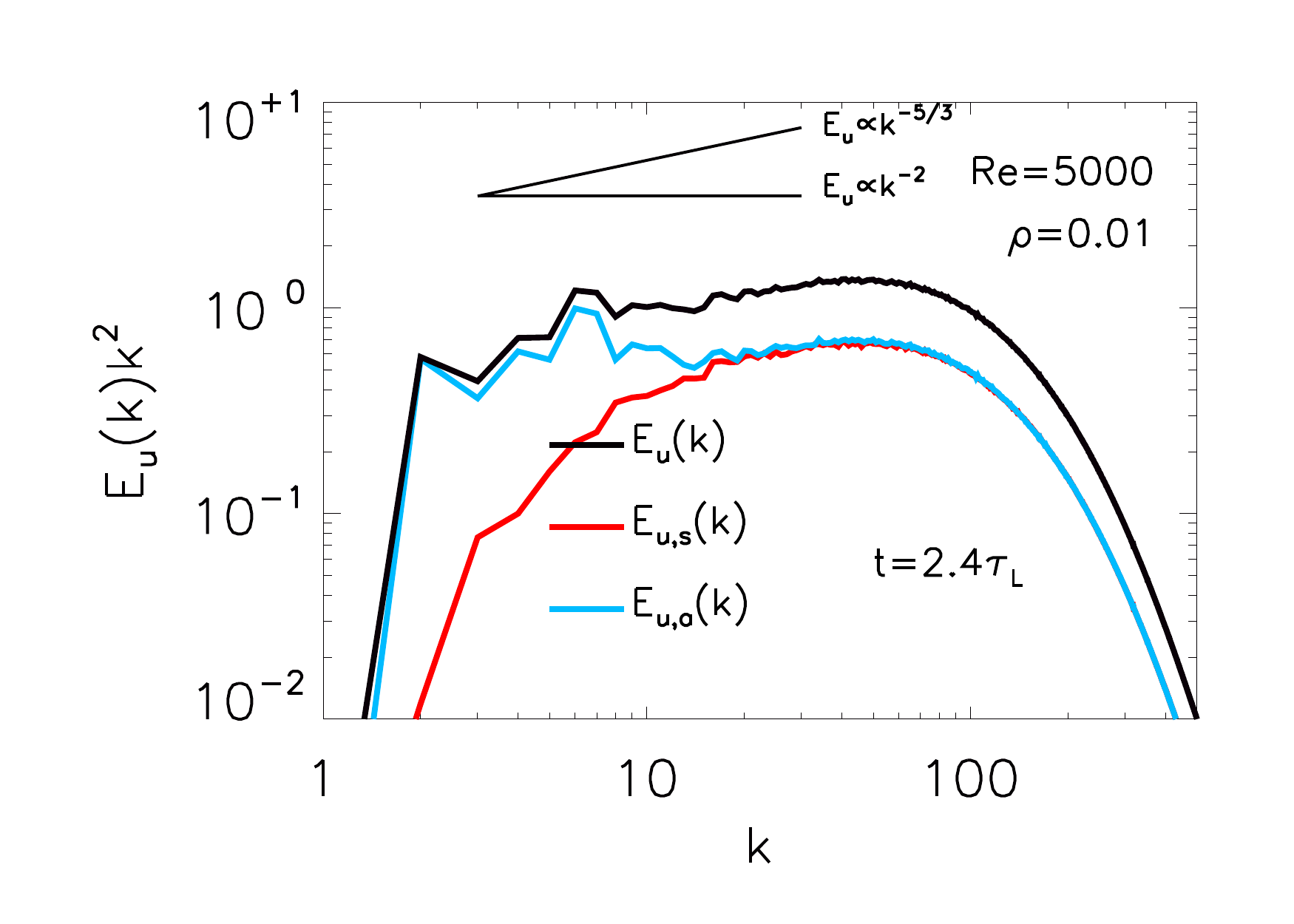}
   \caption{}
  \end{subfigure}
  \caption{(Color online) (a) Magnetic energy and (b) kinetic energy spectra for $\rho=0.01$ and $Re=5000$ at the second peak of total dissipation rate $t/\tau_L=2.4$.}
  \label{fig:SpecE2b}
 \end{figure}

The case with $\rho=0.01$ and $Re=5000$ is more insightful since both the energy and the enstrophy/current density of the perturbation are significantly smaller than the TG initial conditions. The spectra for this case are shown for the two total dissipation peaks %why? 2 peaks only exist for Ohmic dissipation but not for the viscous one!
in Figs. \ref{fig:SpecE2a} and \ref{fig:SpecE2b} at times $t/\tau_L=1.7$ and $t/\tau_L=2.4$, respectively. Both magnetic and kinetic energy spectra reach equipartition between the symmetric and
the asymmetric part of the flow at small scales but not at large scales. 
In particular, at the first total dissipation peak $E_{b,s} \sim E_{b,a}$ and $E_{u,s} \sim E_{u,a}$ only for wavenumbers $k > 30$ (see Fig. \ref{fig:SpecE2a}). The slopes of the compensated full energy spectra $k^2 E_b$ and $k^2 E_u$ are positive but less than the $k^{-5/3}$ scaling. Note that for the symmetric part of the magnetic field $k^2 E_{b,s} \sim const$. Therefore, $k^{-2}$ is still a good scaling for $E_{b,s}$ implying that the change in the slope of $E_b$ is due to the symmetry breaking part of the flow.

In the second dissipation peak more scales have reached equipartition %??? Not really!
between the symmetric and the asymmetric part of the flow (see Fig. \ref{fig:SpecE2b}).
The power law of the full magnetic energy spectrum remains between $k^{-2}$ and
$k^{-5/3}$ while the full kinetic energy spectrum is closer to $k^{-5/3}$. 
%It is worth noticing in the compensated velocity spectra that the the small scales where equipartition has been reached have a greater slope than in the large scales. %??? What do you want to say?
%
%The velocity field at the peak of the viscous dissipation
%has a positive slope. A linear fit gives a value close to the kolmogorov prediction $k^{-5/3}$. Note
%however that that the fluctuations are stronger than the magnetic field case. 
% 
% elapisw
%{\bf We conjecture that this intermediate
%value of the exponent is due to the surprising still small value of the Reynolds number. What 
%we expect for even larger values of $Re$ is that the large scales will show a $k^{-2}$ energy spectrum
%the reflects the presence of the current/vortex sheets while the small scales will recover the 
%true turbulent spectrum  ($k^{3/2}$ or $k^{-5/3}$) and are related to the turbulent fluctuations 
%observed in figure \ref{collorfigs}.
%Because we do not have enough inertial range do distinguish the two slopes the observed slope
%appears as an intermediate value.}
%

\begin{figure}[!ht]
  \begin{subfigure}{8cm}
   \includegraphics[width=\textwidth]{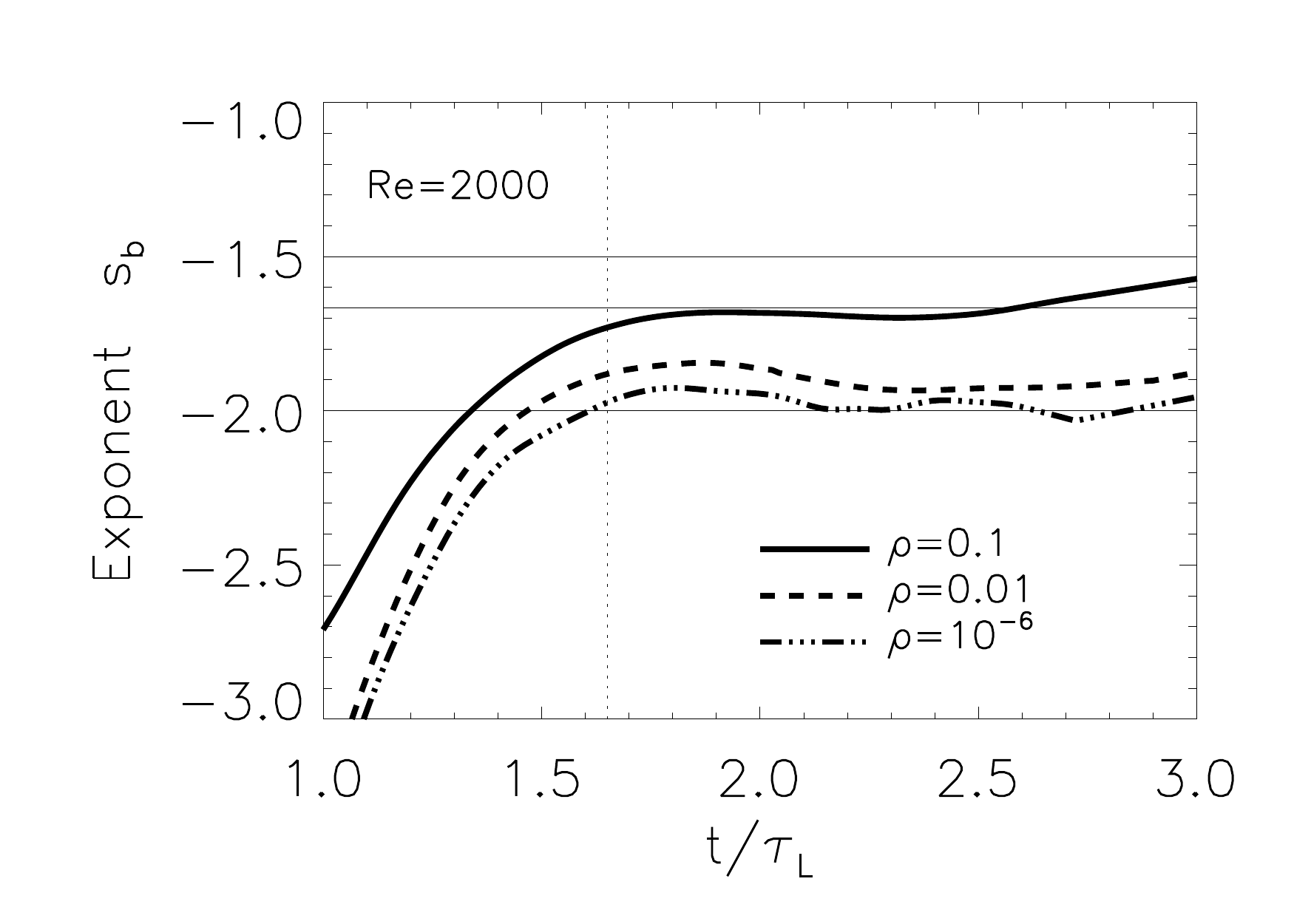}
   \caption{}
  \end{subfigure}
  \begin{subfigure}{8cm}
   \includegraphics[width=\textwidth]{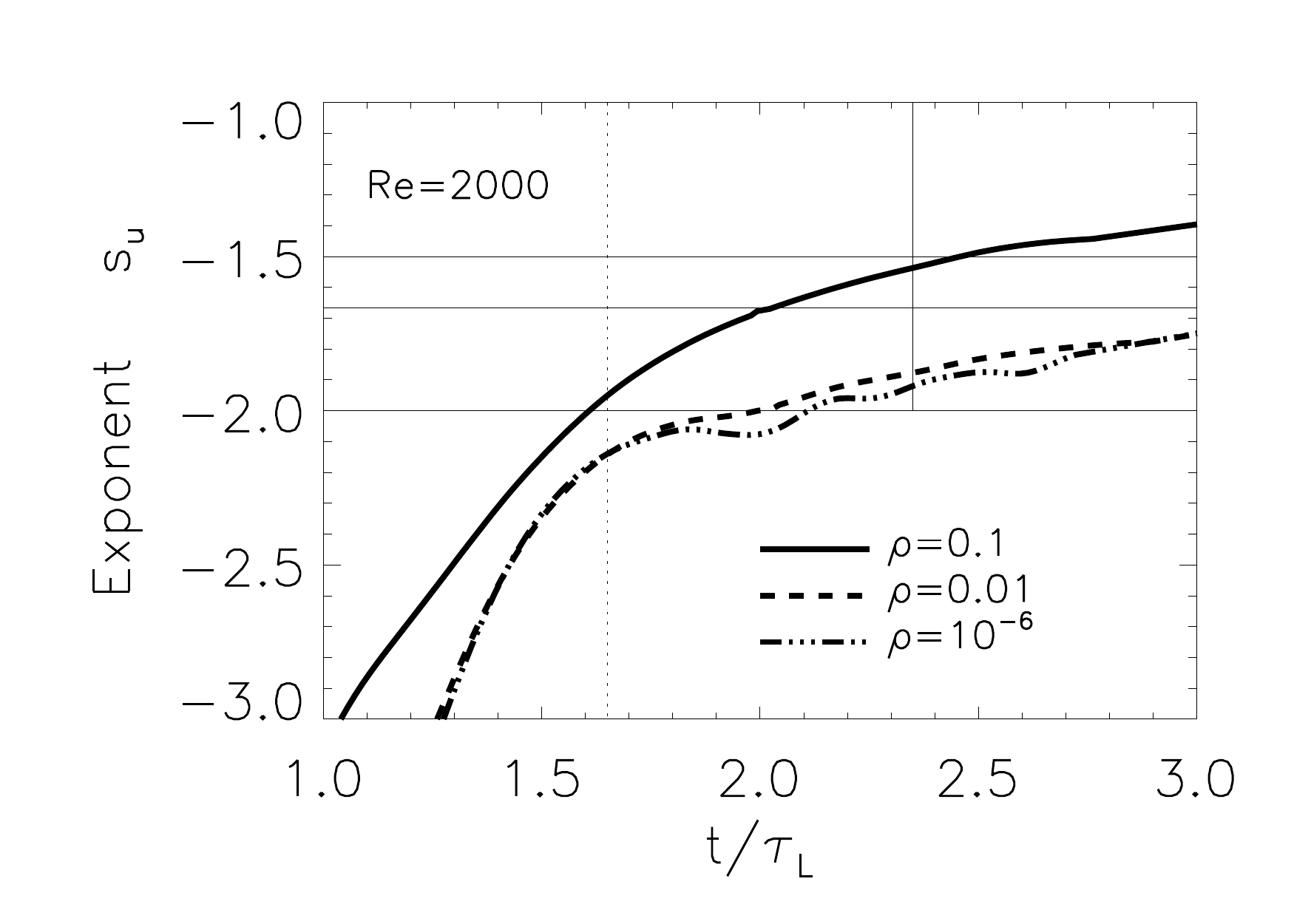}
   \caption{}
  \end{subfigure}
  \caption{Scaling exponent of (a) the magnetic energy and (b) the kinetic energy spectrum
           for different $\rho$ and $Re=2000$.}
  \label{fig:SlopeE}
 \end{figure}

\begin{figure}[!ht]
  \begin{subfigure}{8cm}
   \includegraphics[width=\textwidth]{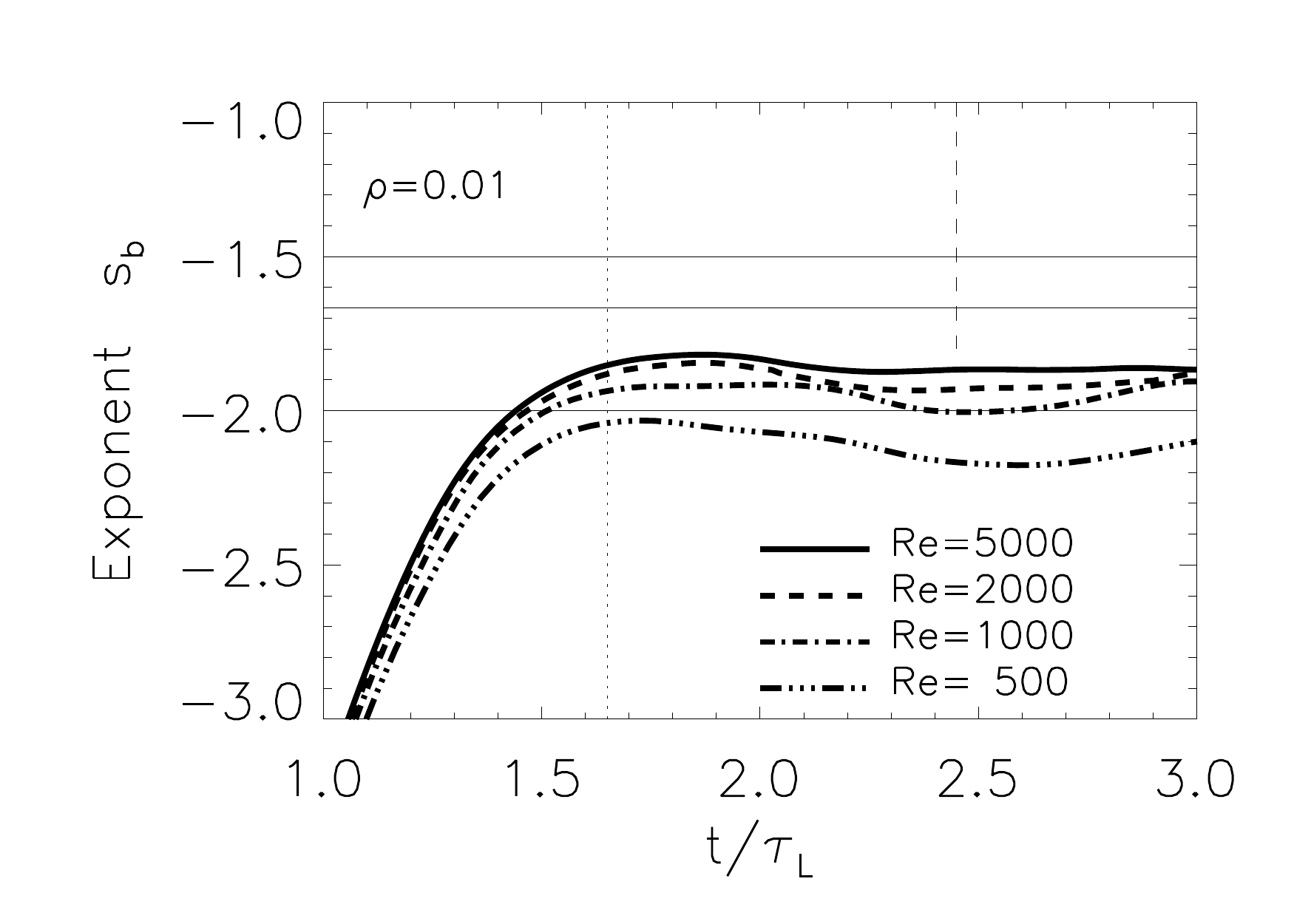}
   \caption{}
  \end{subfigure}
  \begin{subfigure}{8cm}
   \includegraphics[width=\textwidth]{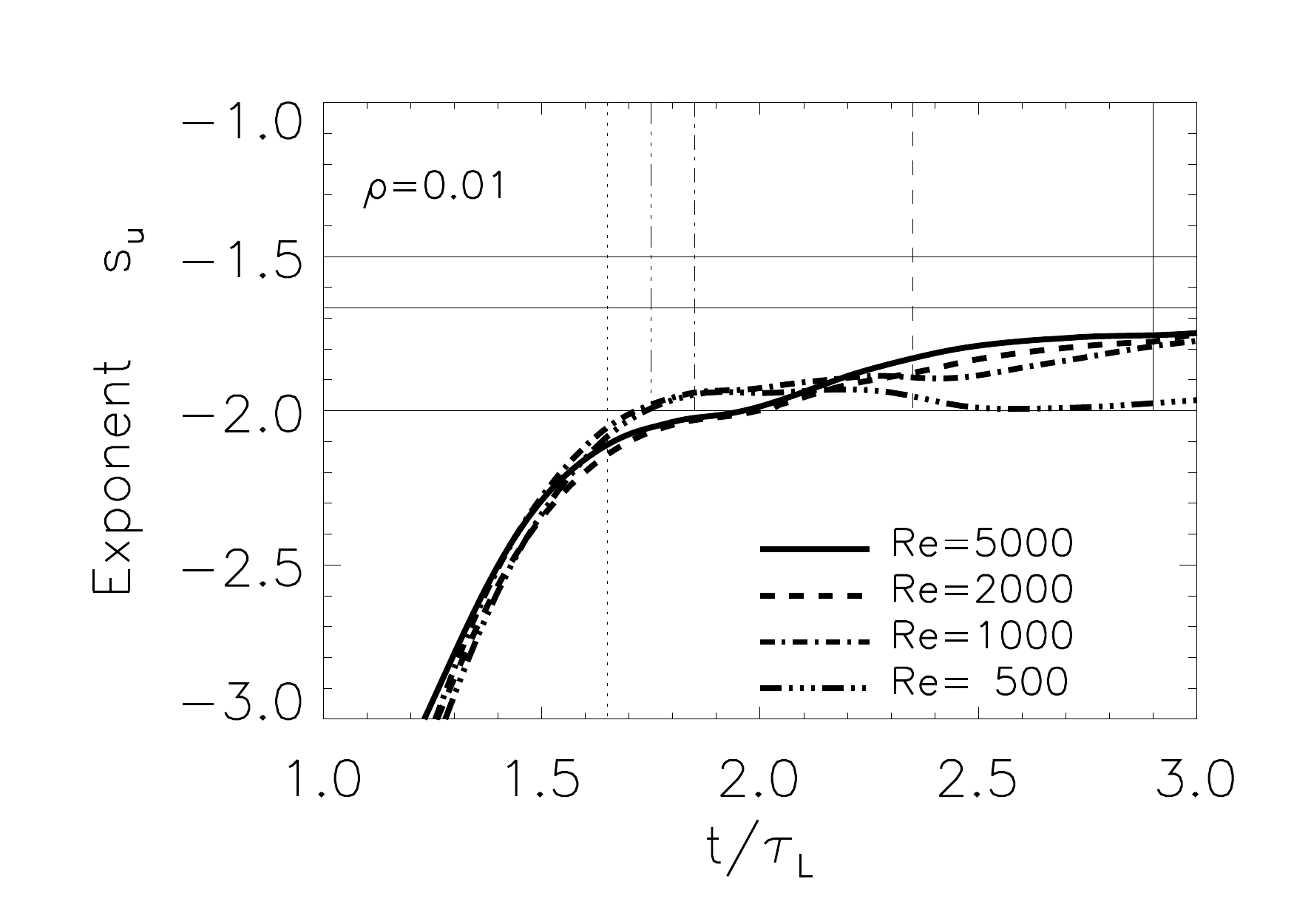}
   \caption{}
  \end{subfigure}
  \caption{Scaling exponent of (a) the magnetic energy and (b) the kinetic energy spectrum for different $Re$ and $\rho=0.01$}
  \label{fig:SlopeR}
 \end{figure}

The evolution of the scaling exponents for the magnetic energy ($E_b \sim k^{s_b}$) and kinetic energy spectra ($E_u \sim k^{s_u}$) for various cases of Table \ref{tbl:dnsparam} are presented in Figs. \ref{fig:SlopeE} and \ref{fig:SlopeR}. 
The scaling exponents were obtained using a linear fit on the energy spectra between wavenumbers $4 < k < 20$ for the runs with $Re=1000$, $4 < k < 30$ for the runs with $Re=2000$ and $4 < k < 40$ for the run with $Re=5000$. %???We should maybe justify in a better way the inertial range and the way to do this is to plot the spectra against k\eta or k\lambda (see also Beresnyak on this issue).
We note that measured exponents in this way are sensitive in the choice of the fitting range especially away from $t_{peak}$. However, the objective here is to show the time evolution of the exponents and not the precise value. Our choices were based on Figs. \ref{fig:SpecE1} to \ref{fig:SpecE2b} as the most reasonable to our opinion. The effect of the perturbation amplitude on the spectral exponents $s_b$ and $s_u$ is shown in Fig. \ref{fig:SlopeE}. 
The strongly perturbed case with $\rho=0.1$ deviates at early times from the other weakly perturbed cases. It saturates to a value close to $-5/3$ for $s_b$ while no clear saturation can be observed for $s_u$. This indicates that Reynolds number is not high enough for a clear scaling of the kinetic energy. Even in the unperturbed case $E_u$ does not have a clear power law scaling (see \cite{da13b}).
The weakly perturbed cases for the magnetic field reach a value close to $s_b = -2$ at the peak of total dissipation and saturate close to this value, % although slightly higher for the magnetic field. 
whereas $s_u$ is close to $-2$ at the total dissipation peak %viscous or Ohmic?
and it drifts to higher values as time progresses.

Figure \ref{fig:SlopeR} focuses on the $\rho=0.01$ case where the highest $Re$ was obtained.
Although $s_u$ is close to $-2$ at the peak of the total dissipation ($t/\tau_L \simeq 1.7$), it increases with time and becomes closer to $-5/3$ as $Re$ increases at the peak of the viscous dissipation (see Fig. \ref{fig:SlopeR}b).
The scaling exponent of the magnetic energy spectrum saturates to a value between $-2$ and $-5/3$. It is worth noting that although the slope at the first peak of Ohmic dissipation ($t/\tau_L \simeq 1.7$) seems to have saturated as a function of $Re$ at the second peak of Ohmic dissipation ($t/\tau_L \simeq 2.4$) the slope appears to still increase with $Re$.

\begin{figure}[!ht]
  \includegraphics[width=8cm]{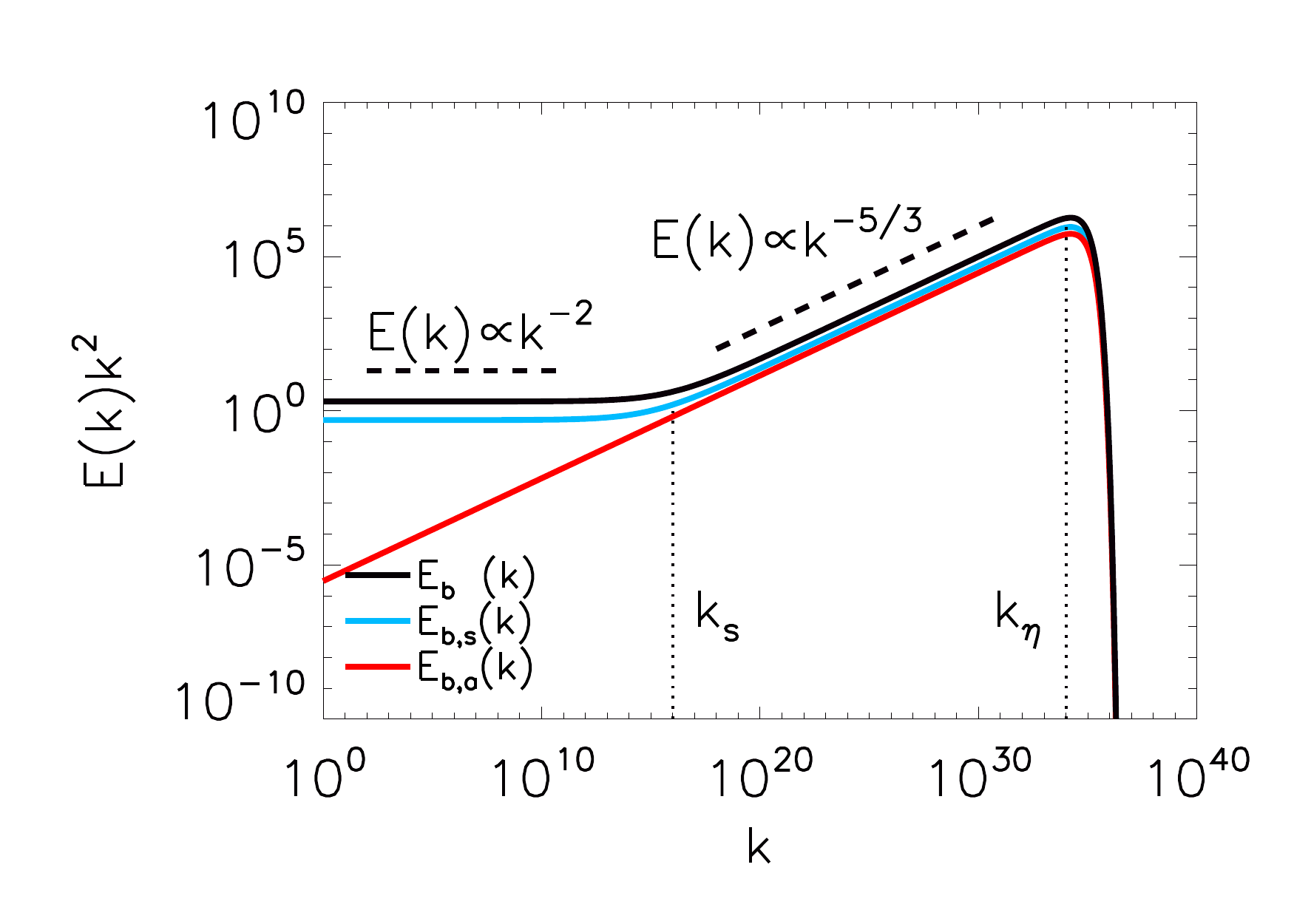}
  \caption{Sketch of an idealised energy spectrum at very high Reynolds number.}
  \label{fig:toyspec}
 \end{figure}

We conjecture that this intermediate value of the exponent, i.e. $-2 < s_b < -5/3$, is a finite Reynolds number effect. We expect that as $Re$ increases the small scales that break the symmetries will start forming a strong turbulence scaling (i.e. $k^{-5/3}$ or $k^{-3/2}$). The spectrum in these scales is related to the turbulent fluctuations observed in Fig. \ref{fig:colorfigs}.
As $Re$ will increase, more small scale turbulent fluctuations will be excited and consequently the range of validity of this power law scaling will increase. Ultimately, this will be the dominant spectrum as $Re \rightarrow \infty$.
This is what we try to depict in Fig. \ref{fig:toyspec} by presenting schematically the expected energy spectra at $Re \gg 1$. For the large scales where the symmetries are not broken, we expect the $k^{-2}$ energy spectrum, which reflects the presence of the strong current/vortex sheets, to persist. The wavenumber $k_s$, that depends on the amplitude of the perturbation, determines the transition point between the $k^{-2}$ and the $k^{-5/3}$ spectrum. If the symmetries are broken at all scales at $Re \gg 1$, then the $-2$ power law spectrum could vanish.
At the present resolutions we do not have enough inertial range to distinguish between different power laws, i.e. one for the small and one for the large wavenumbers. As a result, the scaling exponent that is observed appears as an intermediate value.

%%%%%%%%%%%%%%%%%%%%%%%%%%%%%%%%%%%%%%%%%%%%%
\section{\label{sec:end} Conclusions}
%%%%%%%%%%%%%%%%%%%%%%%%%%%%%%%%%%%%%%%%%%%%%

In this work we have studied one of the proposed initial conditions in \cite{leeetal10} 
for a freely decaying MHD flow. These initial conditions in the absence of any 
perturbation lead to the formation of strong magnetic shear layers that
result in a $k^{-2}$ energy spectrum \cite{da13b} different than the more commonly 
obtained spectra $k^{-5/3}$ and $k^{-3/2}$ for random initial conditions. 
Here, we investigated whether this behaviour persist when the initial conditions weakly deviate
from the ones proposed in \cite{leeetal10} and break the involved Taylor-Green symmetries by adding a small perturbation.

We demonstrated that a sufficiently small perturbation evolves passively and it grows at a rate that increases with Reynolds number. In particular, it was shown that the energy ratio scales as $E_a/E_s \sim Re$  the dissipation ratio scales as $\epsilon_a/\epsilon_s \sim Re^{3/2}$ at the peak of the total dissipation rate. Therefore, for any finite amplitude perturbation, no matter how small it is, there is a high enough Reynolds number for which the perturbation will grow enough at the peak of the total dissipation resulting to a non-linear feedback in the flow and subsequently break the TG symmetries.

For strong perturbations of amplitude $\rho=0.1$ we showed that the TG symmetries break. Turbulent small scales appear both near the strong shearing regions but also in the bulk of the $[0,\pi]^3$ boxes. 
%indicating that the perturbation was large enough. 
These new small scale features change the slope of the energy spectrum from $k^{-2}$ to the classical turbulence spectrum, i.e $k^{-5/3}$ or $k^{-3/2}$ power law scaling.

For the smaller amplitude perturbation $\rho=0.01$ the initially passive asymmetric part of the flow grows to an amplitude that can play a non-linear role in the MHD equations at large $Re$. A new dissipation peak appears as a result of the non-linear evolution of the instability. The strong shearing regions bend and turbulent structures appear at the edge of the current sheet causing this new peak. The scaling exponent of the energy spectrum is clearly larger than $-2$ but still smaller than $-5/3$. This intermediate value of the exponent appears because at the examined Reynolds numbers the small scales have broken the symmetries and are approaching the strong turbulence scaling, while the large scales still exhibit the $k^{-2}$ scaling. Therefore, the measured exponent appears to take an intermediate value. We argue that the strong turbulence scaling (i.e. $k^{-5/3}$ or $k^{-3/2}$) will dominate at higher $Re$.
 
The above results suggest that unless the TG symmetries are satisfied exactly in periodic boxes they will break at sufficiently large $Re$ and the strong turbulence scaling of the spectrum will be recovered. Thus, the system will return to a universal behaviour up to the distinction between the $k^{-5/3}$ and the $k^{-3/2}$ that we could not resolve here.

%There are, however, various issues that still need to be considered and understood. Here we considered a periodic box for which the initial conditions satisfied \textcolor{cyan}{(almost)} the TG symmetries. Another way of seeing the results by \cite{leeetal10} is \textcolor{red}{as simulations in a finite box with free-slip, inductive boundary conditions,} %???
%for which the TG symmetries are satisfied by identity. In this case, the present analysis does not apply and the $k^{-2}$ spectrum still persists. Note, however, that the $k^{-2}$ spectrum results from strong shear layers that are formed at the boundaries \cite{da13b} and \textcolor{red}{close to the boundaries we do not expect a universal behaviour.} %???

Another very important issue is why and under what conditions these large current sheets, which lead to the transient $k^{-2}$ energy spectra form. %Do these structures persist if the flow is continuously forced? 
Even though current sheets form spontaneously in MHD \cite{parker94}, in this case the spanwise length was the size of the fundamental box and their amplitude was strong enough to dominate the energy spectrum something that is not typically observed in random MHD turbulent flows. It is crucial in observations to distinguish the $k^{-2}$ spectra that manifest due to discontinuities in the magnetic field and those due to weak turbulence. 
%Do they manifest due to discontinuities in the magnetic field or due to weak turbulence? 
Note that these mechanisms are distinctly different even though the energy spectra display the same scaling. Some of these questions are going to be addressed in our future work.

\begin{acknowledgements}
V.D. acknowledges the financial support from EU-funded Marie Curie Actions--Intra-European Fellowships (FP7-PEOPLE-2011-IEF, MHDTURB, Project No. 299973). The computations were performed using the HPC resources from GENCI-TGCC-CURIE (Project No. x2013056421) and PRACE-FZJ-JUQUEEN (Project name PRA068).
\end{acknowledgements}

\bibliography{references}
\end{document}